\documentclass[10pt]{article}

\usepackage{amssymb}
\usepackage{color}
\usepackage{epsfig,amssymb,amsfonts,amsmath,graphicx,dsfont,cite,xfrac}
\usepackage{authblk}
\usepackage{subcaption}
\definecolor{mygray}{gray}{0.5}

\usepackage{cite}
\usepackage[colorlinks=true,linkcolor=blue,citecolor=red]{hyperref}

\newcommand{\be}{\begin{equation}}
\newcommand{\ee}{\end{equation}}
\newcommand{\bea}{\begin{eqnarray}}
\newcommand{\eea}{\end{eqnarray}}

\title{Bessel-Gauss beams of arbitrary integer order: propagation profile, coherence properties and quality factor}

\author[1]{S. Cruz~y~Cruz}
\author[2]{Z. Gress}
\author[3]{P. Jim\'enez-Mac\'ias}
\author[3]{O. Rosas-Ortiz}

\affil[1]{\small Instituto Polit\'ecnico Nacional, UPIITA, Av. IPN 2580, C.P. 07340, M\'exico City, Mexico}

\affil[2]{\small Universidad Aut\'onoma del Estado de Hidalgo, Ciudad del Conocimiento, Hidalgo, Mexico}

\affil[3]{\small Physics Department, Cinvestav, AP 14-740, 07000 M\'exico City, Mexico}

\date{}
\begin{document}

\maketitle

\begin{abstract}
We present a novel approach to generate Bessel-Gauss modes of arbitrary integer order and well-defined optical angular momentum in a gradient index medium of transverse parabolic profile. The propagation and coherence properties, as well as the quality factor, are studied using algebraic techniques that are widely used in quantum mechanics. It is found that imposing the well-defined optical angular momentum condition, the Lie group $SU(1,1)$ comes to light as a characteristic symmetry of the Bessel-Gauss beams.
\end{abstract}


\section{Introduction}

The orbital angular momentum of light, more specifically, the orbital angular momentum about the axis of propagation \cite{Bar17}, has remained a lively and growing research topic over the last 30 years (see for example \cite{Bar17,Yao11,Bar17b}). The paper that gave rise to this activity \cite{All92} communicated the idea that laser beams could carry orbital angular momentum, and that this momentum would have measurable effects in the laboratory. Very early, the observation of angular momentum transfer from vortex beams to absorptive particles was reported \cite{He95}, and the orbital angular momentum of light went from assumption to reality, forever changing the way we understand light. This fact arouses immediate interest in the study of the subject  \cite{All00,Pad00,Bar02,Cha18}, specifically in the description, production and detection of modes with well-defined orbital angular momentum \cite{Baz92,Bei93,Arl97,Ngc13,For16,Cru16,Cru20,Gre23}. This property of light is also known as optical angular momentum and goes by the acronym OAM (although there is no universal agreement on what the `O' stands for \cite{Bar17}, as it can mean both orbital and optical). 

OAM beams find natural applications in optical communications \cite{Wil17} and the transfer of information \cite{Rus17}, but are also useful in the quantum domain since OAM can be entangled \cite{Kre17}. Their ability to induce mechanical torque in absorptive systems makes them a critical ingredient in optical micro-manipulation, micro-machining, and optical tweezers \cite{Yan21}. 

Notably, the focusing and collimating properties of parabolic media allow propagation of vortex beams \cite{Cru20,Kot13,Pet16,Cru17,Wu20}. Indeed, the modal delay in the multiplexing of signals is greatly reduced using parabolic optical fibers \cite{Put21} since they tolerate more than two modes having the same propagation constant (effective refractive index).  

Among the most studied vortex beams, the Laguerre-Gauss, Bessel, and Bessel-Gauss modes stand out because their mathematical structure is well known, with which relatively simple theoretical models of their behavior are built, and because their practical realization seems viable in the laboratory \cite{All92,Pad00,Sie86,Dur87a,Dur87b,Gor87,Bag96,Bor97,Li04,Sto23}.

Mathematically, Laguerre-Gauss modes constitute an orthonormal set that is very convenient to span the entire space of paraxial modes. There exist diverse techniques for their production in practice, including spiral plates, computer-generated diffractive gratings, mode converters and spacial light modulators, among others \cite{Yao11}. 

Bessel modes, on the other hand, are non-diffractive and self-healing beams that preserve their transversal structure and reconstruct themselves after passing an obstacle in propagation events. In particular, the zero-order Bessel mode was found to be a solution of the Helmholtz equation in free space \cite{Dur87a}. Indeed, it is a continuous superposition of plane waves for which the wave numbers form a cone around the optical axis. The non-diffraction phenomenon arises because the wave vectors have the same longitudinal component, so the plane wave components undergo the same phase shift and identical intensity distributions are produced as the beam propagates along the optical axis.

However, the Bessel modes are not square integrable, so their production would require an infinite amount of energy in practice. This limitation has been faced with the production of light beams whose intensity distribution acquires the Bessel profile in a finite circular area but which vanishes elsewhere in the transverse plane \cite{Dur87b}. These beams retain some of the most important propagation properties that characterize Bessel beams, at least within a finite distance in the longitudinal direction. For example, they self-replicate by paraxial transformations that can be generated by either free-space propagation, thin lensing, parabolic media, or pure magnifiers \cite{Wol88}. In any case, although the production of Bessel beams can be described analytically, their experimental realization is far from simple.

Looking for analytical alternatives to the Bessel modes, it has already been proposed that the field amplitude profile can be factorized as a Bessel function and a Gaussian envelope. The Fresnel diffraction integral then led to the zero-order Bessel-Gauss mode in the paraxial approximation \cite{Gor87},  a result verified in the laboratory by means of laser cavities \cite{Ueh89}. On the other hand, within a bidirectional travel plane wave scheme, Bessel-Gauss modes of arbitrary integer order can be studied in terms of the exact solutions of the wave equation \cite{Ove91}. They can also be constructed as superpositions of Gaussian modes \cite{Gor87}. In the laboratory, ring arrays of Gaussian beams with wave vectors forming a cylinder or a cone around the propagation axis produce modified and generalized Bessel-Gauss modes, respectively \cite{Bag96}. Due to their non-diffractive properties, the Bessel-Gauss modes allow the design of secure communication protocols \cite{Wan18,Wan23} and are more resilient than the Laguerre-Gauss modes under high-level turbulence conditions \cite{Dos16}.

In this work we study Bessel-Gauss modes of arbitrary integer order, with well-defined OAM, propagating in a gradient index medium with a parabolic inhomogeneity in the transverse direction. Our approach is algebraic and translates notions that are quite natural in quantum mechanics to the field of optics. 

The most striking feature of our approach is to find that not only the profile and the way the BG modes propagate along the $z$-axis, but also their quality is determined by the underlying symmetry of well-defined OAM, which is linked to the Lie group $SU(1,1)$. 

In Section~\ref{Sec2} we analyze the paraxial equation for transverse parabolic media. We show that the well-defined angular momentum condition naturally leads to the solution space being interconnected by the generators of the $su(1,1)$ Lie algebra. Section~\ref{Sec3} is addressed to show that the BG modes are generalized $su(1,1)$ coherent states and explore their coherence properties and quality factor in algebraic terms. A discussion of our results is given in Section~\ref{Sec4}

At the end of the paper we add an appendix with detailed calculations where it is shown that the subspaces of well-defined OAM are irreducible representation spaces for the Lie algebra $su(1,1)$.

\section{Paraxial wave equation for parabolic media}
\label{Sec2}

The paraxial wave equation 
\begin{equation}
\left[- \frac{1}{2k_0^2 n_0} \left( \frac{\partial^2}{\partial x^2} + \frac{\partial^2}{\partial y^2} \right) + n_0 \frac{\Omega^2}{2} r^2 \right] U  =\frac{i}{k_0}  \frac{\partial}{\partial z} U, \qquad \Omega^2 r^2 \ll 1,
\label{phe1}
\end{equation}
describes the $z$-propagation of electromagnetic waves through weakly inhomogeneous media with  refractive index $n^2(r) = n_0^2 \left( 1 -  \Omega^2 r^2 \right)$ \cite{Cru17,Gre17,Gre19,Cru20}. The quantities $n_0 \geq 1$ and $\Omega \geq 0$ represent the refractive index at the optical axis and the strength of the medium inhomogeneity, respectively. In turn, $r \equiv \Vert \boldsymbol{r} \Vert = \sqrt{x^2+y^2}$ denotes the norm of the position-vector $\boldsymbol{r} \in \mathbb R^2$, which is transverse to the propagation axis. From now on  we write $\boldsymbol{r}$ in the polar form, with $\rho \geq0$ and $\theta \in [-\pi, \pi)$.

Equation~(\ref{phe1}) holds if one assumes that the electric field is of the form $\boldsymbol{E}(\boldsymbol{r}, z) = \boldsymbol{E_0} e^{ i(k_0 n_0z-\omega t)} U (\boldsymbol{r},z)$, with $\boldsymbol{E_0}$ the field polarization, $k_0 = \tfrac{\omega}{c}$ the wave number in free space, and $\omega= 2\pi \nu$ the angular frequency. For $\Omega \neq 0$, the complex amplitude $U (\boldsymbol{r},z)$ represents a field mode that is guided within the inhomogeneous medium $n^2(r)$. The limit $\Omega \rightarrow 0$ leads to the homogeneous case $n^2(r)  = n_0^2$. At such a limit, the Rayleigh range $z_R = \tfrac12 k_0n_0w_0^2$ usually denotes the distance from the focal plane ($z=z_0$) over which a Gaussian beam increases its cross-sectional area by a factor of two. In this work we follow \cite{Gre19,Cru20} and set $\Omega z_R=1$ to consider guided modes with constant beam width $w_0$. 

\subsection{Space of solutions (stationary and guided LG modes)}

It may be shown \cite{Cru20} that the functions
\begin{equation}
U_{\ell}^p (\boldsymbol{r}, z) = \frac{e^{i\ell \theta}}{\sqrt{2\pi}}
\exp\left[-i \left( \tfrac{z-z_0}{z_R} \right) \beta_{\vert \ell \vert}^p \right] \Phi_{\vert \ell \vert}^p \left(\rho \right), \quad \ell \in \mathbb Z, \quad p=0,1,2,\ldots,
 \label{ugen}
\end{equation}
are square-integrable solutions of the paraxial wave equation (\ref{phe1}) for $\Omega z_R=1$, where
\be
\beta_{\vert \ell \vert}^p = \vert \ell \vert + 2p + 1
\label{prop}
\ee
stands for the {\em propagation constant}, and 
\begin{equation}
\Phi_{\vert \ell \vert}^p (\rho) =  \frac{2 (-1)^p}{w_0} \sqrt{\frac{\Gamma(p + 1)}{\Gamma(\vert \ell \vert + p + 1)}}  \left[ \frac{\sqrt{2}\rho}{w_0}\right]^{\vert \ell \vert}L_p^{(\vert \ell \vert)} \left(\frac{2\rho^2}{w^2_0}\right)  \exp \left(-\frac{\rho^2}{w^2_0}  \right)
\label{phi}
\end{equation}
is the mode amplitude. In the above expressions $z_0$ defines the origin of the longitudinal coordinate, $L_n^{(\alpha)}(x)$ is the associated Laguerre polynomial of degree $n$ and order $\alpha$ \cite{Olv10}, and $w_0$ stands for the (constant) beam width. 

The mode amplitude (\ref{phi}) is defined on the plane transverse to the propagation direction, as a function of the polar radial coordinate $\rho$, and is weighted by the Gaussian distribution of standard deviation $\sigma = w_0/\sqrt{2}$. That is, $\Phi_{\vert \ell \vert}^p (\rho)$ represents the electric field amplitude of a {\em stationary Laguerre-Gauss mode}. 

The longitudinal and polar phases appearing in Eq.~(\ref{ugen}) supply dynamical properties to the  solutions $U_{\ell}^p (\boldsymbol{r}, z)$; they refer to the propagation of the beam and the helicity of the mode, respectively. The propagation occurs in the direction in which the variable $z$ increases. In turn, the orientation of the helicity is based on the polar phase increments: if these are counterclockwise the helicity is right-handed, otherwise (clockwise) it is left-handed. The former is characterized by $\ell >0$, and the latter by $\ell <0$. Modes with null orbital angular momentum ($\ell = 0$) do not exhibit helicity. 

Note that the helically phased modes $U_{\ell}^p$ and $U_{-\ell}^p$ have the same amplitude and longitudinal phase since neither the function (\ref{phi}) nor the propagation constant (\ref{prop}) depend on $\operatorname{sgn}(\ell)$.

From now on the functions $U_{\ell}^p (\boldsymbol{r}, z)$ introduced in (\ref{ugen}) will be referred to as {\em guided Laguerre-Gauss modes} (LG modes for short). 

In applications where a Gaussian distribution is the desired profile, the beam propagation factor $\mathcal{M}^2$ (also called beam quality factor) is the most important feature describing the quality of the beam. It was introduced to characterize laser beams by comparing their beam parameter product (waist-radius~$\times$~divergence) with that of a Gaussian beam \cite{Sie90,Sie93}. As a parameter, it determines how close a given light field is to an ideal Gaussian beam \cite{Sal07,Bel94}. 

For the LG modes  (\ref{ugen}), the factor $\mathcal{M}^2$ coincides with the propagation constant (\ref{prop}). The lowest value $\beta_0^0 =1$ refers to the fundamental LG mode $U_0^0$, which represents a Gaussian beam. Higher order modes $U_{\ell}^{p}$ may be grouped according to the nonnegative integer $n = \vert \ell \vert + 2p$, for which we write $\beta_{\vert \ell \vert}^p \equiv \beta_{(n)}=n+1$. In this form, $\ell$ and $p$ can be combined in $n+1$ different ways to get the same value
\be
\beta_{\vert \ell \vert}^p = \underbrace{\beta_{\vert -n \vert}^0= \beta_{\vert -n+2 \vert}^1 = \cdots = \beta_{n-2}^1 = \beta_n^0}_{n+1 \operatorname{terms}}\equiv \beta_{(n)} = n+1 ; \quad n= \vert \ell \vert + 2p = \operatorname{fixed}.
\nonumber
\ee
That is, there are $n+1$ modes $U_{\ell}^p$ with exactly the same propagation constant $\beta_{\vert \ell \vert}^p = \beta_{(n)}$. In other words, $\beta_{\vert \ell \vert}^p$ is $(n+1)$-fold degenerate for $n=\vert \ell \vert + 2p = \operatorname{fixed}$. 

\begin{figure}[h!]
\centering 
\includegraphics[width=.7\textwidth]{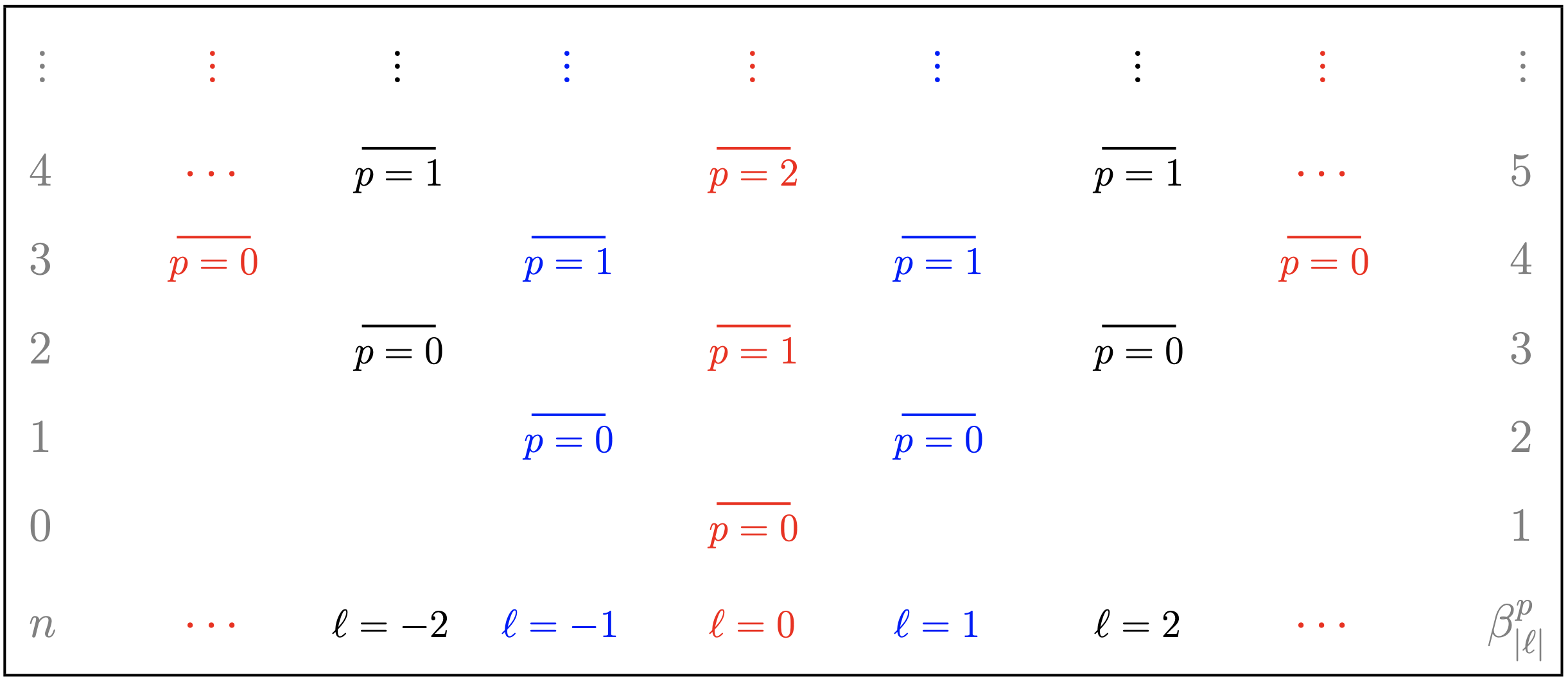}

\caption{\footnotesize  The propagation constant (\ref{prop}) is (n + 1)-fold degenerate for $n= \vert \ell \vert + 2p =\operatorname{fixed}$. In such a case we write $\beta_{\vert \ell \vert}^p \equiv \beta_{(n)}=n+1$ (horizontal arrays). For $\ell = \operatorname{fixed}$, the admissible values of $\beta_{\vert \ell \vert}^p$ are equidistant, with steps of two units (vertical arrays). In general, the entire set $\{\beta_{\vert \ell \vert}^p \}$ may be counted by a one-to-one correspondence with the energy spectrum of a two-dimensional harmonic oscillator in the quantum domain.
}
\label{energy}
\end{figure}

For practical purposes, one may identify the ensemble $\{ \beta_{(n)} \}$ with the energy eigenvalues of a two-dimensional quantum harmonic oscillator (2D quantum oscillator for short), see Figure~\ref{energy}.

The set $\{ U_{\ell}^p \}$ is orthonormal and compete according to the inner product 
\be
\int_{-\pi}^{\pi} \int_0^\infty U_\ell^{\, p *} (\boldsymbol{r}, z) U_l^q  (\boldsymbol{r}, z) \rho d\rho d\theta =  \delta_{\ell,l} \delta_{p,q},
\label{orto}
\ee
where $u^*$ stands for the complex-conjugate of $u \in \mathbb C$. Hence, the LG modes (\ref{ugen}) form an orthonormal basis for the solution space $\mathcal{H}$ of the paraxial wave equation (\ref{phe1}). We formally write
\[
\mathcal{H}= \operatorname{span} \{ U_{\ell}^p (\boldsymbol{r},z); \; \ell \in \mathbb Z , \, p=0,1,2,\ldots\}.
\]
Given (\ref{orto}), each of the functions $U_{\ell}^p (\boldsymbol{r}, z)$ represents a collimated beam that carries finite transverse optical power as it propagates along the $z$-axis \cite{Cru20}.   

\subsection{Subspaces with a well-defined optical angular momentum}

It is convenient to classify the LG modes into {\em hierarchies of well-defined optical angular momentum} $\ell \in \mathbb Z$, which are defined as the solution subspaces
\[
\mathcal{H} \supset \mathcal{H}_{\ell} = \operatorname{span} \{ U_{\ell}^p (\boldsymbol{r},z)  ; \, \ell = \operatorname{fixed}, \, p=0,1,2,\ldots \}, \quad \ell \in \mathbb Z.
\]
The hierarchies $\mathcal{H}_{\ell}$ are infinite-dimensional and satisfy $\mathcal{H} = \bigoplus_{\ell} \mathcal{H}_{\ell}$. They are characterized by a denumerable set of propagation constants $\beta_{\vert \ell \vert}^p$ that are equidistant, with steps of two units $\beta_{\vert \ell \vert}^{p+1} - \beta_{\vert \ell \vert}^p=2$, see Figure~\ref{energy}. 

Considering that LG modes are helically phased, we may write $\mathcal{H}^-= \bigoplus_{\ell < 0} \mathcal{H}_{\ell}$, $\mathcal{H}^+= \bigoplus_{\ell >0} \mathcal{H}_{\ell}$, and $\mathcal{H}^0 = \mathcal{H}_{\ell =0}$ for the solution subspaces with clockwise, counterclockwise and null helicity, respectively. Then, the entire space of solutions is also written as $\mathcal{H}  = \mathcal{H}^- \bigoplus \mathcal{H}^0 \bigoplus \mathcal{H}^+$.

Hereafter we work on the elements of the hierarchy $\mathcal{H}_{\ell}$. Our interest is addressed to manipulate the coherence properties of electromagnetic modes with a well-defined optical angular momentum. With this in mind, the association of the ensemble $\{ \beta_{\vert \ell \vert}^p \}$ with the energy spectrum of the 2D quantum oscillator is very useful.

By translating the algebraic structure of contemporary quantum mechanics to study the optical system we are dealing with, we have a beautiful tool at hand for designing LG-mode superpositions on demand. In particular, we look for the optimization of the coherence properties of wave packets with a well-defined optical angular moment. 

We have successfully carried out this translation, details are given in Appendix~\ref{ApA}. There, a series of ladder operators is constructed such that a certain LG mode $U_{\ell}^p (\boldsymbol{r},z)$ is connected to another $U_l^q (\boldsymbol{r},z)$, where $p$ and $q$, and $\ell$ and $l$, are different in general.

The most striking feature of the algebraic formalism developed in Appendix~\ref{ApA} is that the symmetry underlying the hierarchy $\mathcal{H}_{\ell}$ is associated with the $su(1,1)$ Lie algebra:
\be
[ \mathcal{L}^-_{\ell}, \mathcal{L}^+_{\ell}] = 2 \mathcal{L}_{\ell}, \quad [ \mathcal{L}_{\ell}, \mathcal{L}^{\pm}_{\ell} ] = \pm \mathcal{L}^{\pm}_{\ell},
\nonumber
\ee
the generators of which are defined as follows
\begin{equation}
\mathcal{L}_{\ell}^- = e^{i 2\frac{(z-z_0)}{z_R}} \frac1{\sqrt{\rho}} \; L_{\ell}^- \sqrt{\rho}, 
\qquad \mathcal{L}_{\ell}^+ = e^{-i 2\frac{(z-z_0)}{z_R}} \frac1{\sqrt{\rho}} \; L_{\ell}^+ \sqrt{\rho}, \quad 
\mathcal{L}_{\ell} =  \frac1{\sqrt{\rho}} \; L_{\ell} \sqrt{\rho}.
\nonumber
\end{equation}
Here, $L^{\pm}_{\ell}$ and $L_{\ell}$ are the second-order differential operators
\be
L^{\pm}_{\ell} = \frac12 \left( \frac{w_0}{2} \right)^2 \left[ \frac{\partial^2}{\partial \rho^2} - \frac{ (\ell + \frac12)(\ell - \frac12)}{\rho^2} \right] \mp \frac12 \rho \frac{\partial}{\partial \rho}
+ \frac12 \frac{\rho^2}{w_0^2} \mp \frac14,
\nonumber
\ee
and
\be
L_{\ell} = \frac12 \left( \frac{w_0}{2} \right)^2 \left[ - \frac{\partial^2}{\partial \rho^2} + \frac{ (\ell + \frac12)(\ell - \frac12)}{\rho^2} \right] + \frac12 \frac{\rho^2}{w_0^2}.
\nonumber
\ee
Note that $L_{\ell}$ coincides with the radial part of the energy observable of a 2D quantum oscillator in position-representation.

The LG mode $U_{\ell}^p (\boldsymbol{r}, z)$ is eigenfunction of $\mathcal{L}_{\ell}$ with eigenvalue $\lambda_p = \frac12 \beta_{\vert \ell \vert}^p = \frac{\vert \ell \vert}2 + p + \frac12$ (to simplify the notation, we have made the dependence of $\lambda_p$ on $\ell$ implicit). The ensemble $\{ \lambda_p \}$ mimics the energy spectrum (shifted by $\frac{\vert \ell \vert}2$) of the one-dimensional harmonic oscillator in quantum mechanics, see Table~\ref{table1}. In turn, $\mathcal{L}_{\ell}^{\pm}$ are ladder operators of the LG modes. That is, the action of $\mathcal{L}_{\ell}^{\pm}$ on $U_{\ell}^p (\boldsymbol{r}, z)$ produces an eigenfunction of $\mathcal{L}_{\ell}$ with eigenvalue $\lambda_p \pm 1 = \lambda_{p \pm 1}$.

\begin{table}[h!]
\begin{center}
\caption{\footnotesize The Laguerre-Gauss modes $U_{\ell}^p (\boldsymbol{r},z)$ are eigenfunctions of $\mathcal{L}_{\ell}$ with eigenvalue $\lambda_p = \frac{\vert \ell \vert}2 + p + \frac12$ (it is shown only the information for the first three values of $p$ and $\vert \ell \vert$). The number $p$ of nodes admitted by the mode along the $\rho$-axis defines both, $\lambda_p$ and the propagation constant $\beta_{\vert \ell \vert}^p = 2 \lambda_p$, and thus determines the way the mode propagates along the $z$-axis.
\label{table1}}
\begin{tabular}{ccccc}
\hline
$\lambda_p$ & $\ell =0$ & $ \ell  = \pm 1$ & $ \ell  =\pm 2$ & $ \ell  = \pm3$\\
 \hline
$\lambda_0$ & 1/2 &1 & 3/2 & 2\\[.5ex]
$\lambda_1$ & 3/2 &2 & 5/2 & 3\\[.5ex]
$\lambda_2$ & 5/2 & 3 & 7/2 & 4\\[.5ex]
$\lambda_3$ & 7/2 &4 & 9/2 & 5\\[.5ex]
 \hline
\end{tabular}
\end{center}
\end{table}

The nonnegative integer $p$ corresponds to the number of nodes admitted by the LG mode along the $\rho$-axis. That is, $p$ refers to the behavior of the mode on the plane transverse to the propagation direction. Although the latter is stationary, the value of $p$ defines the propagation constant (eigenvalue) $\beta_{\vert \ell \vert}^p = 2 \lambda_p$ since $\ell = \operatorname{const}$ in $\mathcal{H}_{\ell}$ and thus determines the way the mode propagates along the $z$-axis. Therefore, it is appropriate to find a way to determine $p$ in the basis elements of $\mathcal{H}_{\ell}$.

In Appendix~\ref{ApA} we have introduced a differential operator $\hat n_p$, called the {\em harmonic-number} (number for short), which acts on the transverse (stationary) part of the LG modes and returns the corresponding number of nodes, as requested. In this form, according to the number drawn by $\hat n_p$, the eigenvalues (propagation constants) $\lambda_p = \frac12 \beta_{\vert \ell \vert}^p$ define the fundamental harmonic ($p=0$) as well as the higher harmonics ($p \geq 1$) in any superposition of LG modes. The role that $\hat n_p$ plays in the description of light beams with well-defined optical angular momentum is of fundamental importance, as we are going to see.

\section{Guided Bessel-Gauss modes as generalized coherent states}
\label{Sec3}

Once we have incorporated the algebraic structure of quantum mechanics in the description of light beams with well-defined optical angular momentum, it is acceptable to translate some quantum mechanical concepts to the arena of electromagnetic waves. Of course, the latter are still defined by the Maxwell theory, regardless the algebraic techniques used in their analysis seem more natural in quantum theory.

Of particular interest, the notion of coherence introduced by Glauber in quantum optics \cite{Gla07} is intrinsically algebraic, so it has been generalized to encompass systems other than light (for a recent review, see \cite{Ros19}).

In what follows, we adhere to the more general notion of a coherent state: it is a linear superposition that exhibits some specific properties that are determined by the `user' based on the phenomenology under study or on theoretical arguments \cite{Ros19}. With this in mind, we look for superpositions of LG modes with a well-defined optical angular momentum.

According to Barut and Girardello \cite{Bar71}, the normalized solution of the eigenvalue equation
\be
\mathcal{L}_{\ell}^- \, V_{\ell} (\boldsymbol{r}, z; \xi) = \xi \, V_{\ell} (\boldsymbol{r},z; \xi), \quad \xi = \tau e^{-i\phi}, \quad \tau \geq 0, \quad \phi \in [-\pi, \pi),
\label{barut}
\ee
defines a coherent state for the Lie algebra we are dealing with.

To solve Eq.~(\ref{barut}) we first express $V_{\ell} (\boldsymbol{r},z; \xi)$ in terms of the basis of $\mathcal{H}_{\ell}$,
\be
V_{\ell} (\boldsymbol{r},z; \xi) = \sum_{p=0}^{\infty} c_p U_{\ell}^p (\boldsymbol{r}, z), \quad c_p \in \mathbb{C}.
\nonumber
\ee
The coefficients $c_p$ are defined by the action of the ladder operator $\mathcal{L}_{\ell}^-$ on the LG modes $U_{\ell}^p (\boldsymbol{r}, z)$, so the precise form of $V_{\ell} (\boldsymbol{r},z; \xi)$ is directly associated with the symmetry underlying the hierarchy $\mathcal{H}_{\ell}$, which is characterized by the Lie algebra $su(1,1)$. The explicit calculation yields
\be
V_{\ell} (\boldsymbol{r},z; \xi) = 
\left[\frac{2 \tau^{\vert \ell \vert}}{\pi w_0^2 I_{\vert \ell \vert} (2 \tau)}\right]^{1/2}
 \left[ \frac{\sqrt{2}\rho}{w_0}\right]^{\vert \ell \vert}
e^{i \ell \theta}  e^{-\frac{i}{z_R} (\vert \ell \vert + 1) (z-z_0)} 
 e^{-\frac{\rho^2}{w^2_0} }
\Lambda_{\ell} (\rho,z,\xi),
\nonumber
\ee
where $I_{\nu}$ stands for the modified-Bessel function \cite{Olv10}, and $\Lambda_{\ell} (\rho,z,\xi)$ is the superposition
\begin{equation}
\Lambda_{\ell} (\rho,z,\xi) = 
\sum_{p=0}^{\infty} \frac{\left(\tau e^{-i \phi}e^{-i \frac{2}{z_r}(z-z_0) } e^{i\pi}\right)^p}{(\vert \ell \vert + p)!}  L_p^{(\vert \ell \vert)} \left(\frac{2\rho^2}{w^2_0}\right).
\nonumber
\end{equation}
Further simplification is achieved by using the series and connection formulae \cite{Gra07,Olv10} 
\be
J_\nu(2 \sqrt{uv}) e^v (uv)^{-\frac{\nu}2}
= \sum_{k=0}^\infty \frac{v^k}{(k+\nu)!} L_k^{(\nu)} (u), \qquad J_{\nu} (u e^{\pm i \pi/2}) = e^{\pm i\nu \pi/2} I_{\nu} (u),
\label{formulas}
\ee
where $J_\nu$ denotes the Bessel function of the first kind. We finally write
\be
\begin{array}{c}
V_{\ell} (\boldsymbol{r},z; \xi) = 
\exp \left\{ i \left[\ell \theta + \frac{\vert \ell \vert}{2} \phi   - \frac{(z-z_0)}{z_R} + \tau \sin \left(\phi + \frac{2}{z_R}(z-z_0) \right) \right] \right\}\\[2ex]
\hskip20ex \times \; \left[\frac{2}{\pi w_0^2 I_{\vert \ell \vert} (2 \tau)}\right]^{1/2}
 \exp \left[-\frac{\rho^2}{w_0^2} - \tau \cos \left(\phi + \frac{2}{z_R}(z-z_0) \right) \right] \\[2ex]
\hskip20ex \times \;   I_{\vert \ell \vert} \left( \frac{2 \sqrt{2\tau}}{w_0} \rho e^{-\frac{i}{2} \phi } e^{- \frac{i}{z_R}(z-z_0) }
\right). 
\label{cs}
\end{array}
\ee
Hereafter the functions (\ref{cs}) will be referred to as {\em Bessel-Gauss coherent states} (BG coherent states or BG modes for short). We also simplify notation by making $z_0=0$. 

Next, we clarify the role played by the real and imaginary parts of the eigenvalue $\xi = \tau e^{-i\phi}$ in the description of the BG modes.

$\bullet$ {\bf Fundamental Gaussian mode.} Using the properties of the modified-Bessel function $I_{\nu}$ \cite{Gra07,Olv10} one has
\be
\lim_{\xi \rightarrow 0} V_{\ell} (\boldsymbol{r},z; \xi) = \left\{
\begin{array}{cc}
\sqrt{\frac{2}{\pi w_0^2}}  \exp \left(-i\frac{z}{z_R} \right) \exp\left( -\frac{\rho^2}{w_0^2} \right), & \ell=0\\[2ex]
0, & \ell \neq 0
\end{array}
\right. .
\label{vlim}
\ee
That is, for $\xi =0$ there is only one regular solution to the Barut-Girardello equation (\ref{barut}), given by the fundamental Gaussian mode $V_{\ell =0} (\boldsymbol{r},z; 0) = U_{\ell =0}^{p=0}  (\boldsymbol{r},z)$, which is a coherent state by definition.

$\bullet$ {\bf Modified-Bessel and Bessel profiles.} For $\xi \neq 0$ the functions $V_{\ell} (\boldsymbol{r},z; \xi)$ consist of a modified-Bessel function $I_{\vert \ell \vert}$ modulated by a Gaussian distribution of standard deviation $\sigma = w_0/{\sqrt 2}$. The latter accelerates the radial decay of the beam. Depending on the eigenvalue-phase $\phi$, the function $I_{\vert \ell \vert}$ is interchanged by $J_{\vert \ell \vert}$, and vice versa (the same occurs at specific points along the propagation axis, see details below). 

For example, making $\phi=0$, the profile of $V_{\ell} (\boldsymbol{r},z; \tau)$ is determined by the modified-Bessel function $I_{\vert \ell \vert}$ in (\ref{cs}). However, from the connection formula included in Eq.~(\ref{formulas}), we see that $\phi = - \pi$ replaces $I_{\vert \ell \vert}$ with the Bessel function $J_{\vert \ell \vert}$ in (\ref{cs}). Thus, we have
\be
\begin{array}{c}
V_{\ell} (\boldsymbol{r},z; - \tau) = \left[\frac{2}{\pi w_0^2 I_{\vert \ell \vert} (2 \tau)}\right]^{1/2}
\exp \left\{ i \left[\ell \theta  - \frac{z}{z_R} - \tau \sin \left( \frac{2z}{z_R} \right) \right] \right\}\\[2ex]
\hskip20ex \times \;  
\exp \left[-\frac{\rho^2}{w_0^2} +\tau \cos \left(\frac{2z}{z_R} \right) \right]  
J_{\vert \ell \vert} \left( \frac{2 \sqrt{2\tau}}{w_0} \rho e^{- i \frac{z}{z_R} }
\right).
\end{array}
\label{cs1}
\ee
This result shows that the phase $\phi$ of the complex-eigenvalue $\xi \neq 0$ characterizes the Bessel profile of our coherent states.

\begin{figure}[h!]

\centering
\subfloat[][Expected value of the number operator $\hat n_p$]{\includegraphics[width=.35\textwidth]{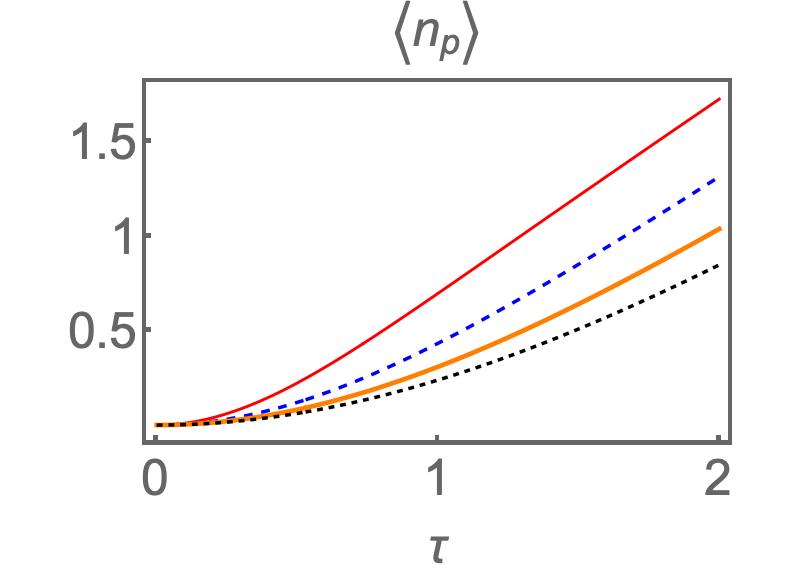}} 
\hskip5ex
\subfloat[][Expected value of the OAM operator $\mathcal{L}_{\ell}$]{\includegraphics[width=.35\textwidth]{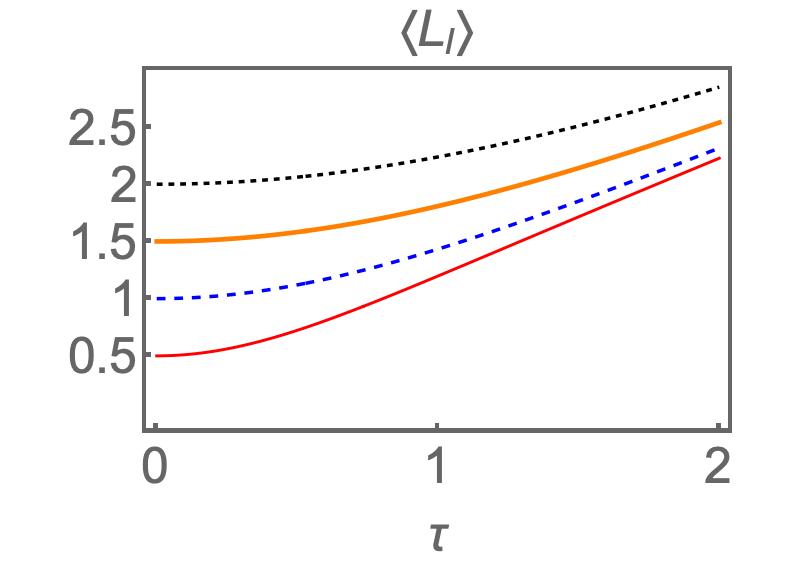}}

\caption{\footnotesize The quality of the Bessel-Gauss modes $V_{\ell} (\boldsymbol{r},z; \xi)$ can be manipulated in terms of the parameter $\vert \xi \vert = \tau$. The harmonics $\langle \hat n_p \rangle =p$ occur at smaller values of $\tau$ for smaller values of $\vert \ell \vert$ (\textbf{a}). The fundamental harmonic ($p=0$) is only permitted for $\ell =0$. Therefore, the lowest eigenvalue $\langle \mathcal{L}_{\ell} \rangle = \lambda_0 = \frac12 (\vert \ell \vert +1)$ never occurs in higher BG modes (\textbf{b}). The BG modes are as close to the ideal Gaussian beam as the expected value  $\langle \mathcal{L}_{\ell} \rangle$ is close to $\lambda_0$, which occurs in the vicinity of $\tau = 0$. The color-code is as follows: $\ell =0$ (continuous, narrow-red), $\ell =1$ (dashed, blue), $\ell =2$ (continuous, wide-orange), and $\ell =3$ (dotted, black).
}
\label{power}
\end{figure}

$\bullet$ {\bf Contribution of the $p$-th harmonic.}  The most likely eigenvalue $\lambda_p$ occurring in the superposition $V_{\ell} (\boldsymbol{r},z; \xi)$ is determined by the expectation value $\langle \mathcal{L}_{\ell} \rangle = \langle \hat n_p \rangle + \frac12 ( \vert \ell \vert + 1)$. The helicity parameter $\ell$ is fixed, so the relevant information is encoded in the expectation value of the number operator $\hat n_p$. The straightforward calculation yields
\be
\langle \hat n_p \rangle = \frac{\tau I_{\vert \ell \vert + 1}(2\tau)}{I_{\vert \ell \vert}(2\tau)}.
\label{energia}
\ee
Figure~\ref{power}(a) shows the behavior of $\langle \hat n_p \rangle$ as a function of $\vert \xi \vert = \tau$. In general, it can be seen that $\langle \hat n_p \rangle$ decreases quadratically with $\tau<1$ but  increases linearly with $\tau>1$. The latter is easily verified from the properties of the modified-Bessel function $I_{\nu}$ \cite{Olv10}, so one has
\be
\langle \hat n_p \rangle_{\tau \ll1} \approx \left( \tfrac{1}{\vert \ell \vert +1} \right) \tau^2, \qquad \langle \hat n_p \rangle_{\tau \gg1} \approx \tau.
\label{cotas}
\ee

In Figure~\ref{power}(a) we see that the fundamental harmonic $\langle \hat n_{p=0} \rangle =0$ occurs at $\tau =0$, which by necessity implies $\ell =0$, see Eq.~(\ref{vlim}). The higher harmonics $\langle \hat n_p \rangle = p \geq 1$ occur at smaller values of $\tau$ for smaller values of $\vert \ell \vert$.

On the other hand, Figure~\ref{power}(b) shows that the higher eigenvalues $\langle \mathcal{L}_{\ell} \rangle = \lambda_p$, $p \geq 1$, occur at larger values of $\tau$ for larger values of $\vert \ell \vert$. Remarkably, with exception of the fundamental BG mode, the lowest eigenvalue $\langle \mathcal{L}_{\ell} \rangle = \lambda_0 $ never occurs in higher BG modes since  $\langle \hat n_{p=0} \rangle =0$ is not defined for $\ell \neq 0$.

The latter means that only the fundamental BG mode behaves like a Gaussian beam. In turn, the higher BG modes are as close to the ideal Gaussian beam as the expected value  $\langle \mathcal{L}_{\ell} \rangle$ is close to $\lambda_0$ (see Table~\ref{table1}), which occurs in the vicinity of $\tau = 0$.

Clearly, the modulus of the complex eigenvalue $\xi $ is responsible for the quality of the BG modes. By setting $\tau$ close to zero, we can make the BG modes as Gaussian as the limit $\tau \rightarrow 0$ allows.

\begin{figure}[h!]

\centering
\subfloat[][Fundamental mode contribution]{\includegraphics[width=.32\textwidth]{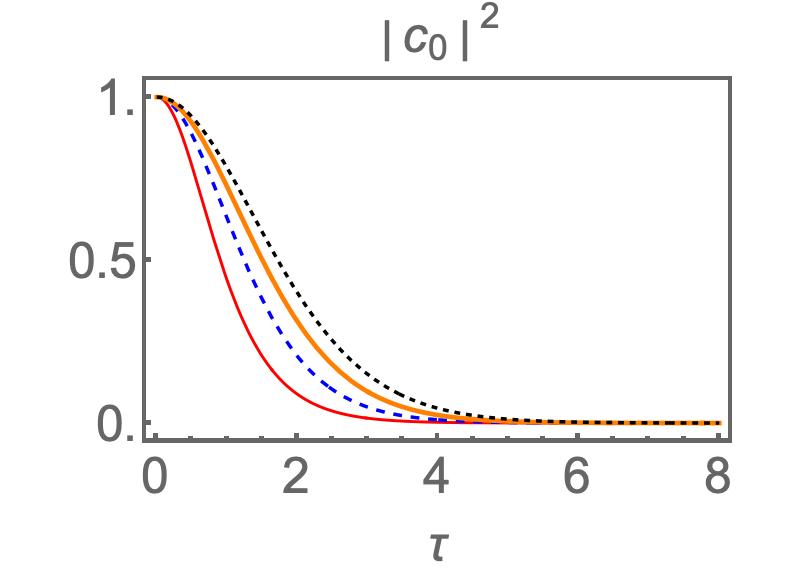}} 
\hskip.5ex
\subfloat[][First mode contribution]{\includegraphics[width=.32\textwidth]{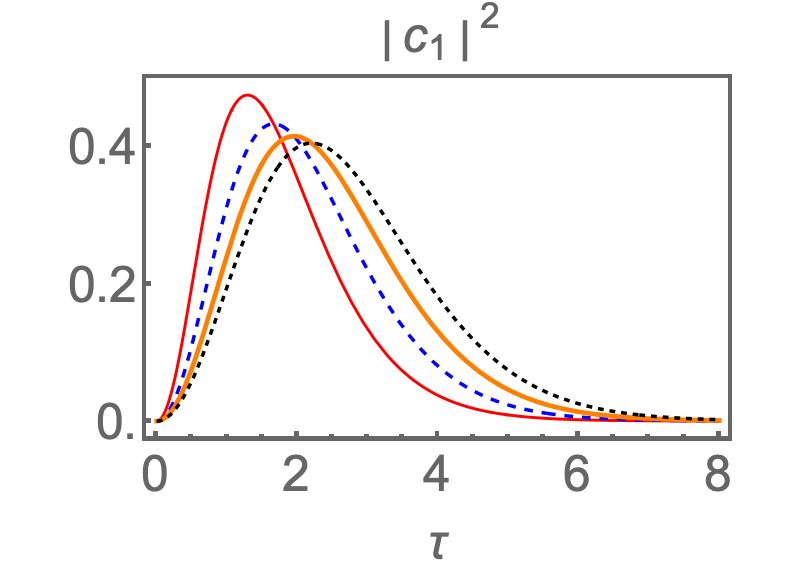}}
\hskip.5ex
\subfloat[][Second mode contribution]{\includegraphics[width=.32\textwidth]{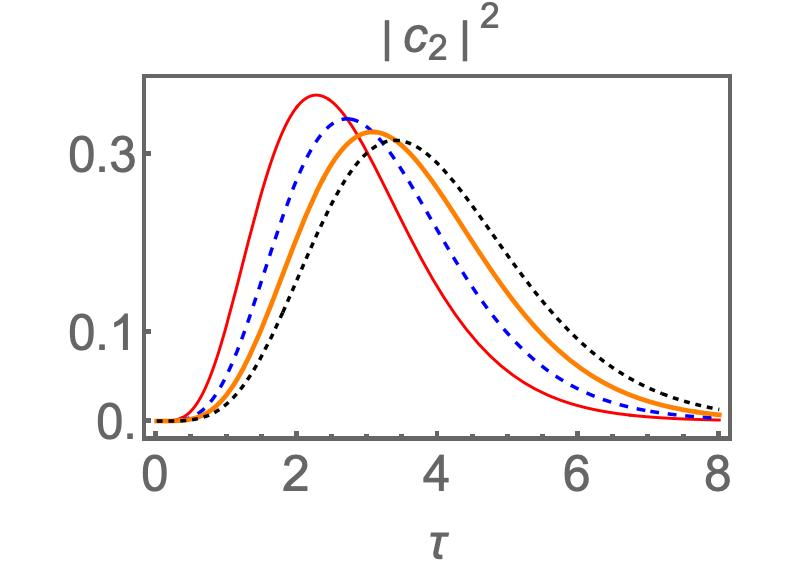}}

\caption{\footnotesize The harmonics labeled with small $p$ contribute more significantly to increasing the quality of the BG mode. By setting $\tau$ close to zero, the main contribution is provided by the fundamental mode (\textbf{a}). The remaining harmonics (figures \textbf{b} and \textbf{c} show $p=1$ and $p=2$) contribute by perturbing the fundamental harmonic in such a way that the BG mode deviates from the ideal Gaussian profile. The color-code is the same as in Figure~\ref{power}.
}
\label{cp}
\end{figure}

The contribution of the $p$-th harmonic to the coherence of the BG mode $V_{\ell} (\boldsymbol{r},z; \xi)$ is more easily visualized using the square modulus of the related coefficient
\[
\vert c_p (\ell, \tau) \vert^2 = \frac{ \tau^{\vert \ell \vert + 2p}}{ I_{\vert \ell \vert} (2 \tau) \Gamma (p+1) \Gamma( \vert \ell \vert + p +1)}.
\]
For small values of $\tau$, the above expression acquires the form
\[
\vert c_p (\ell, \tau) \vert^2 \approx \frac{ \tau^{2p} \, \Gamma (\vert \ell \vert + 1) }{\Gamma (p+1) \Gamma( \vert \ell \vert + p +1)}.
\]
In the vicinity of $\tau = 0$ one has $\tau^0 > \tau^2 > \tau^4 > \cdots$. Therefore, the harmonics labeled with small $p$ contribute more significantly to increasing the quality of the BG mode. This is illustrated in Figure~\ref{cp}. Clearly, $c_0$ provides the largest square modulus for $\tau \approx 0$, so the fundamental LG mode is dominant in any superposition intended to build high-quality BG modes. The contribution of the remaining harmonics can be treated as a disturbance (noise) that deviates the BG mode from the ideal Gaussian profile.

\subsection{Variances and standard deviations}

We now introduce the quadratures
\be
\mathcal{L}_{1; \ell}= \frac{ \mathcal{L}_{\ell}^+ + \mathcal{L}_{\ell}^-}{2}, \qquad \mathcal{L}_{2; \ell}= \frac{ \mathcal{L}_{\ell}^+ - \mathcal{L}_{\ell}^-}{2i},
\nonumber
\ee
which are non-commuting variables $[\mathcal{L}_{1;\ell}, \mathcal{L}_{2;\ell} ] = -i \mathcal{L}_{\ell}$. Using the BG modes, the related variances\footnote{The variance and standard deviation are respectively given by $\mu_{\hat u} = \langle \hat u^2 \rangle - \langle \hat u \rangle^2$ and $\sigma_{\hat u} =\sqrt{\mu_{\hat u}}$.} are equally weighted:
\be
\mu_{\mathcal{L}_{1; \ell}} = \mu_{\mathcal{L}_{2; \ell}} =  \tfrac12 \langle \mathcal{L}_{\ell} \rangle.
\nonumber
\ee
As a consequence, the inequality
 \[
 \sigma_{\mathcal{L}_{1;\ell}} \sigma_{\mathcal{L}_{2;\ell}} \geq \tfrac12 \vert \langle [\mathcal{L}_{1;\ell}, \mathcal{L}_{2;\ell} ] \rangle \vert,
 \] 
named after Robertson \cite{Rob29}, is saturated. 

The latter means that the average errors of $\mathcal{L}_{1;\ell}$ and  $\mathcal{L}_{2; \ell}$ are not only the same, but minimal (it is said that the BG modes are {\em minimum uncertainty states}). Then, it may not be  possible to prepare a state with both distributions, $\sigma_{\mathcal{L}_{1; \ell}} $ and $\sigma_{\mathcal{L}_{2; \ell}}$, sharply concentrated around any concrete value \cite{Ken27}. Therefore, to characterize the BG modes $V_{\ell} (\boldsymbol{r},z; \xi)$ we need to set up another uncertainty relationship, the one that holds for the appropriate observables.

In the formal operator description of paraxial wave optics, the canonical conjugate variables of transverse-position $\boldsymbol{r}$ and transverse-propagation direction $\boldsymbol{p}$ are treated as self-adjoint operators, see for example \cite{Glo69,Sto81}. Using the BG modes, the mean value of such variables vanishes for the on-axis beams, $\langle \boldsymbol{r} \rangle  = \langle \boldsymbol{p} \rangle  =0$. Additionally, the straightforward calculation yields $\langle [ \boldsymbol{r}, \mathbf{p} ] \rangle = \frac2{k_0}$, together with
\be
\mu_{\boldsymbol{r}} =  w^2_0 \left[ \langle \mathcal{L}_{\ell} \rangle  + \tau \cos \left( \tfrac{2z}{z_R} +\phi \right) \right], \qquad 
\mu_{\boldsymbol{p}} = \tfrac{4}{k_0^2 w_0^2} \left[  \langle \mathcal{L}_{\ell} \rangle - \tau \cos \left( \tfrac{2z}{z_R} +\phi \right)  \right].
\label{mus}
\ee
Therefore, the Robertson inequality for $\boldsymbol{r}$ and $\boldsymbol{p}$ gives
\begin{equation}
\sigma_{\boldsymbol{r}}\sigma_{\boldsymbol{p}}= \tfrac{2}{k_0} \left[ \langle \mathcal{L}_{\ell}\rangle^2 -\tau^2 \cos^2 \left(\tfrac{2 z}{z_R} +\phi \right)  \right]^{1/2} \geq \tfrac{1}{k_0}.
\label{sigmas}
\end{equation}
If $\tau \neq 0$, the transverse-spreading $\sigma_{\boldsymbol{r}}\sigma_{\boldsymbol{p}}$ oscillates between two extreme values
\be 
\sigma_{\boldsymbol{r}}\sigma_{\boldsymbol{p}} \vert_{\operatorname{max}} = \tfrac{2}{k_0} \langle \mathcal{L}_{\ell} \rangle, \qquad
\sigma_{\boldsymbol{r}} \sigma_{\boldsymbol{p}} \vert_{\operatorname{min}} = \tfrac{2}{k_0} \sqrt{ \langle \mathcal{L}_{\ell} \rangle^2  -\tau^2}, 
\label{extreme}
\ee
which are periodically attained at $z_n= \tfrac12 [ (n+ \tfrac12) \pi -\phi ] z_R$ and $z_q = \tfrac12 (q\pi -\phi) z_R$, respectively, with $n,q \in \mathbb Z$. 

The above results exhibit the self-focusing and collimation properties of the BG modes,  and show that the corresponding beam is collimated. 

By necessity, if $\tau=0$ then $\ell=0$. In such a case, the transverse-spreading $\sigma_{\boldsymbol{r}}\sigma_{\boldsymbol{p}}$ is constant and acquires the global minimum value 
\be
\sigma_{\boldsymbol{r}}\sigma_{\boldsymbol{p}} \vert_{\operatorname{Gauss}} = k_0^{-1},
\ee 
so it saturates inequality (\ref{sigmas}). The latter because $V_{\ell =0} (\boldsymbol{r},z; \xi=0)$ is the fundamental Gaussian mode $U_0^0  (\boldsymbol{r},z)$, see Eq.~(\ref{vlim}).

\begin{figure}[htb]
\centering
\subfloat[][Transverse-spreading $\sigma_{\boldsymbol{r}}\sigma_{\boldsymbol{p}}$]{\includegraphics[width=.35\textwidth]{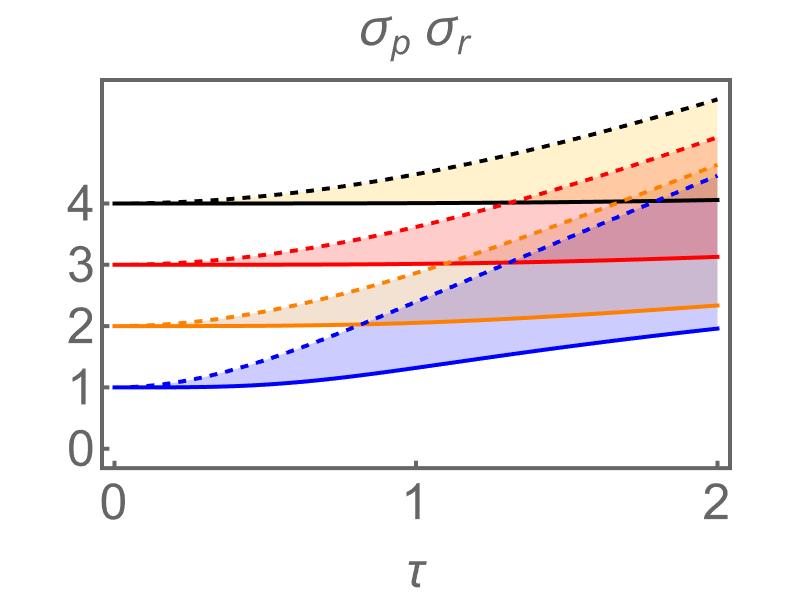}} 
\hskip5ex
\subfloat[][Beam quality factor $M^2_{\ell} = k_0 \sigma_{\boldsymbol{r}}\sigma_{\boldsymbol{p}} \vert_{\operatorname{min}}$]{\includegraphics[width=.35\textwidth]{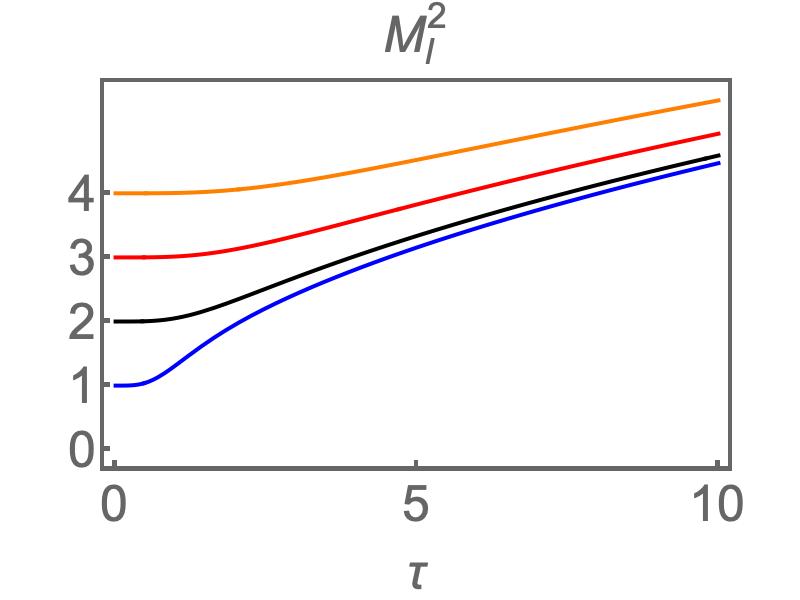}}

\caption{\footnotesize The transverse-spreading $\sigma_{\boldsymbol{r}}\sigma_{\boldsymbol{p}}$ introduced in (\ref{sigmas}) for the Bessel--Gauss coherent states $V_{\ell } (\boldsymbol{r},z; \xi)$ oscillates between the extreme values $\sigma_{\boldsymbol{r}} \sigma_{\boldsymbol{p}} \vert_{\operatorname{max}}$ and $\sigma_{\boldsymbol{r}}\sigma_{\boldsymbol{p}} \vert_{\operatorname{min}}$, dashed and continuous curves, respectively. The areas between these values are shadowed to facilitate the comparison of results (\textbf{a}). The function $\sigma_{\boldsymbol{r}}\sigma_{\boldsymbol{p}} \vert_{\operatorname{max}}$ refers also to the mean energy and the quadrature deviations, compare (\textbf{a}) with Figure~\ref{power}. The minimal value $\sigma_{\boldsymbol{r}}\sigma_{\boldsymbol{p}} \vert_{\operatorname{min}}$ is directly proportional to the $M_{\ell}^2$ factor (\ref{M2}), plotted in (\textbf{b}). In all cases, the curves correspond to  $\ell = 3,2,1,0$ (from top to bottom). Horizontal and vertical axis are in units of $\tau$ and $k_0^{-1}$, respectively. The lowest bound defined by both, the Robertson and the Schr\"odinger inequalities, is attained at $\tau=0$ for $\ell=0$, written $\sigma_{\boldsymbol{r}}\sigma_{\boldsymbol{p}} \vert_{\operatorname{Gauss}} = k_0^{-1}$. 
}
\label{ineq}
\end{figure}

Figure~\ref{ineq}(a) shows a comparison between the extreme values (\ref{extreme}) of the transverse-spreading $\sigma_{\boldsymbol{r}}\sigma_{\boldsymbol{p}}$, as a function of $\tau = \vert \xi \vert$, for $\ell=0,1,2,3$. The smaller $\tau$, the shorter the difference $\sigma_{\boldsymbol{r}}\sigma_{\boldsymbol{p}} \vert_{\operatorname{max}}- \sigma_{\boldsymbol{r}}\sigma_{\boldsymbol{p}} \vert_{\operatorname{min}}$, which confirms that the beam is better collimated for small values of $\tau$. 

\subsection{Beam quality}

We would like to emphasize that $\langle \boldsymbol{r} \rangle = \langle \boldsymbol{p} \rangle =0$ is just a necessary condition (but not sufficient) for $\boldsymbol{r}$ and $\boldsymbol{p}$ to fluctuate totally independent one from another. Indeed, after Schr\"odinger \cite{Sch30} it is currently known that the Robertson inequality includes the skew Hermitian part $[ \boldsymbol{r}, \boldsymbol{p} ]$ of the product $\boldsymbol{rp}$, but it omits the complementary Hermitian part. In fact, taking into account the complete decomposition $\boldsymbol{rp} = \frac12 \{ \boldsymbol{r}, \boldsymbol{p} \}  +\frac12 [ \boldsymbol{r}, \boldsymbol{p} ]$, with $\{\cdot , \cdot \}$ the anticommutator of the involved operators, we should follow Schr\"odinger, who proposed the more precise inequality
\be
\sqrt{ \mu_{\hat u} \mu_{\hat v} -\left[ \tfrac12 \langle \{ \hat u, \hat v \} \rangle - \langle \hat u \rangle \langle \hat v \rangle \right]^2} \geq \tfrac12 \left \vert \langle [\hat u, \hat v] \rangle
\right \vert.
\label{ineq1}
\ee
The term in square brackets under the square-root symbol is known as covariance and measures the joint variability of $\hat u$ and $\hat v$ when non-commutability is taken into account. 

In addition to (\ref{mus}), for the present case one has $\langle \{ \boldsymbol{r}, \mathbf{p}\} \rangle = -\frac4{k_0} \tau \sin (\tfrac{2 z}{z_R}+\phi)$. Then, the Schr\"odinger inequality (\ref{ineq1}) yields  $ \sqrt{ \langle \mathcal{L}_{\ell} \rangle^2  -\tau^2} \geq \frac12$. That is, considering the non commutability of $\boldsymbol{r}$ and $\boldsymbol{p}$, the Robertson inequality  (\ref{sigmas}) must be corrected to only compare $\sigma_{\boldsymbol{r}}\sigma_{\boldsymbol{p}} \vert_{\operatorname{min}}$ with the lowest bound $k_0^{-1}$ (that is, with the transverse-spreading of an ideal Gaussian beam). Taking this conclusion into account, we introduce the function
\be
M^2_{\ell} (\tau) := k_0 \sigma_{\boldsymbol{r}}\sigma_{\boldsymbol{p}} \vert_{\operatorname{min}}= k_0 \sqrt{\mu_{\boldsymbol{r}} \mu_{\mathbf{ p}} - \tfrac14 \langle \{ \boldsymbol{r}, \mathbf{p}\} \rangle^2 } = 2 \sqrt{\langle \mathcal{L}_{\ell} \rangle^2  -\tau^2},
\label{M2}
\ee
which satisfies $M^2_{\ell} (\tau) \geq 1$, see Figure~\ref{ineq}(b). The lowest bound $M^2_0 (0) =1$ corresponds to the fundamental Gaussian mode $V_{\ell =0} (\boldsymbol{r},z; \xi=0) = U_0^0  (\boldsymbol{r},z)$.

The function (\ref{M2}) may be identified with the $\mathcal{M}^2$ factor of light signals \cite{Sal07,Bel94}, which is often used to specify the beam quality \cite{Sie90,Sie93}: the higher the value of $\mathcal{M}^2$, the lower is the beam quality. By definition, the lowest value $\mathcal{M}^2=1$ is assigned to Gaussian beams, which are called diffraction-limited beams.

Indeed, the very last expression of (\ref{M2}) coincides with the $\mathcal{M}^2$ factor obtained in \cite{Bor97} for generic Bessel--Gauss beams. The authors of \cite{Bor97} develop a direct calculation of the second-order moments associated to the intensity distributions at the waist plane and in the far field, denoted $\sigma_0$ and $\sigma_{\infty}$ respectively, to write $\mathcal{M}^2 = 2\pi \sigma_0 \sigma_{\infty}$. Here, we have simplified the derivation of such a result by using algebraic techniques, with no cumbersome calculations involved.

Additionally, the penultimate expression of (\ref{M2}) is commonly used as a definition of the $\mathcal{M}^2$ factor when the first order moments of the canonical variables vanish, as for the case of on-axis paraxial beams. But the calculation of moments is extremely difficult in general, so there are only few examples where it has been possible \cite{Ban10,Mar09}. It is then extremely interesting to provide either models allowing the calculation of higher order moments or innovative approaches addressed to circumvent technical difficulties. However, it is notable that the intimate relation between the beam quality factor $\mathcal{M}^2$ and the Schr\"odinger inequality (\ref{ineq1}) is barely recognized in the literature. Remarkable exceptions are \cite{Dod00}, where it is suggested a link between some invariant quantities and generalized uncertainty relations like the Schr\"odinger one, and \cite{Cru17}, where the $\mathcal{M}^2$ factor of Hermite-Gauss modes is directly related to the Schr\"odinger inequality for $\boldsymbol{r}$~and~$\boldsymbol{p}$. 

In the present work we have exploited the fact that the Schr\"odinger correction to the Robertson inequality (\ref{sigmas}) eliminates the sinusoidal $z$-dependence of the product $\sigma_{\boldsymbol{r}}\sigma_{\boldsymbol{p}}$, and considers only the minimal value $\sigma_{\boldsymbol{r}}\sigma_{\boldsymbol{p}} \vert_{\operatorname{min}}$ to measure the joint variability of $\boldsymbol{r}$ and $\boldsymbol{p}$ when non-commutability is taken into account. It is then relevant to find that the $M_{\ell}^2$ factor (\ref{M2}) coincides with $\sigma_{\boldsymbol{r}}\sigma_{\boldsymbol{p}} \vert_{\operatorname{min}}$.

For small values of $\tau$, the $\mathcal{M}^2$ factor behaves as follows
\be
\left. M^2_{\ell}(\tau)  \right\vert_{\tau \ll 1} \approx \frac{2 \tau^4}{(\vert \ell \vert +1)^3} + \vert \ell \vert +1.
\label{Mlim}
\ee
At the very limit $\tau \rightarrow 0$, the lowest value $M^2_{\ell}(\tau=0) = \vert \ell \vert +1$ is only reached for $\ell =0$. That is, the ideal Gaussian profile $M^2_{\ell =0}(\tau=0) = 1$ occurs for the fundamental mode $V_{\ell =0} = U_{\ell =0}^{p=0}$ only, as we have already mentioned. 

For $\ell \neq 0$, the quality of the BG mode obeys the rule $M^2_{\ell}(\tau) > \vert \ell \vert +1$. How close $M^2_{\ell}(\tau)$ is to its lower bound depends on how close $\tau$ is to zero. 

However, keep in mind that the optical angular momentum $\ell \hbar$ spoils the beam quality: poor beam quality results for large $\vert \ell \vert$, no matter how small $\tau $ is.

Therefore, our previous conclusion is reinforced: the fundamental LG mode is dominant in wave packets leading to high-quality BG modes. The remaining harmonics contribute by producing noise, so the BG mode deviates from the ideal Gaussian profile. The smaller the value of the mean $\langle \hat n_p \rangle$, the better is the quality of the BG mode. High mean harmonic-number means a very large factor $M_{\ell}^2$ for the BG beams.

\subsection{Propagation properties}

\begin{figure}[htb]
\centering
\includegraphics[width=.7\textwidth]{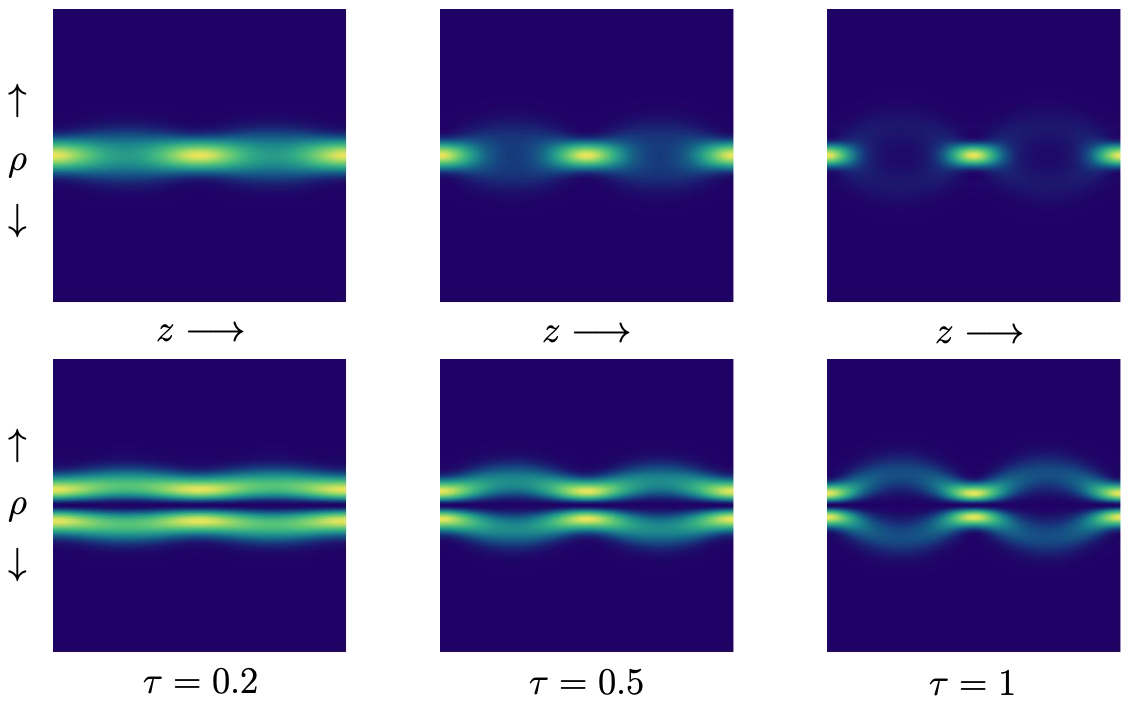}

\caption{\footnotesize Intensity of the BG coherent states $V_{\ell} (\boldsymbol{r},z; \xi)$  introduced in (\ref{cs}), from the longitudinal perspective, for $\xi = -\tau$ and the indicated values of $\tau$. The function $V_{\ell} (\boldsymbol{r},z; \xi)$ is periodic with period $2\pi z_R$. {\bf Collimation:} The beam is better collimated for small values of $\tau$. {\bf Self-focusing:} The configuration at $z=0$, where the radial uncertainty $\sigma_r$ is minimum and the value of $\sigma_p$ as well as intensity are maxima, is recovered at $z_q=q \pi z_R$. In turn, the maximum value of $\sigma_r$ (minimum value of $\sigma_p$) and minimum intensity are attained at $z_n= (n+\tfrac12) \pi z_R$. In both cases the transverse-spreading $\sigma_r \sigma_p$ reaches its minimum value. {\bf Transverse profile:} For $\ell=0$ the intensity is different from zero at $\rho=0$ (upper row), but it cancels at $\rho=0$ for $\vert \ell \vert =1$ (lower row). The behavior for $\vert \ell \vert \geq 2$ is qualitatively equivalent to the case $\vert \ell \vert =1$.
}
\label{propaga}
\end{figure}

Note that although the BG modes (\ref{cs}) are superpositions of LG modes with constant width, they exhibit self-focus since the propagation constants are commensurable, so their initial transversal profile is reproduced periodically along the propagation direction, see Figure~\ref{propaga}. 

The intensity $\vert V_{\ell} (\boldsymbol{r},z; \xi) \vert^2$ is symmetrical with respect to rotations around the propagation axis, and periodical in $z$ with period $\pi z_R$ since $V_{\ell} (\boldsymbol{r},z; \xi)$ is periodical in $z$ with period $2\pi z_R$. For $\ell=0$ and any value of $z$, the intensity is different from zero at $\rho=0$, so it depicts a spot on the transverse plane. However, if $\ell \neq 0$, the intensity cancels at $\rho=0$ and is distributed in annular way on  the transverse plane for any value of $z$. 

The maximum transverse-spreading of the beam is parameterized by $\vert \xi \vert = \tau$. On the other hand, for propagation along the $z$-axis, the following periodical identities are obeyed
\be
\begin{array}{c}
V_{\ell} (\boldsymbol{r}, 2 z_q; \xi) = V_{\ell} (\boldsymbol{r}, 0; \xi), \quad \vert V_{\ell} (\boldsymbol{r},z_q; \xi) \vert^2 = \vert V_{\ell} (\boldsymbol{r}, 0; \xi) \vert^2, 
\\[2ex]
V_{\ell} (\boldsymbol{r}, z_n; \xi) =  (-1)^{n (\vert \ell \vert +1)} V_{\ell} (\boldsymbol{r}, \tfrac{\pi}{2} z_R; \xi), \quad \vert V_{\ell} (\boldsymbol{r}, z_n; \xi) \vert^2 =  \vert V_{\ell} (\boldsymbol{r}, \tfrac{\pi}{2} z_R; \xi) \vert^2, 
\end{array}
\label{iden2}
\ee
where 
\[
z_q = q\pi z_R, \quad z_n = (n+ \tfrac12) \pi z_R, \quad q,n \in \mathbb Z,
\]
and
\be
\begin{array}{c}
V_{\ell} (\boldsymbol{r}, \tfrac{\pi}{2} z_R; \xi) = \left[\frac{2}{\pi w_0^2 I_{\vert \ell \vert} (2 \tau)}\right]^{1/2}
\exp \left\{ i \left[\ell \theta + \frac{\vert \ell \vert}{2} \phi -\tfrac{\pi}{2}(\vert \ell \vert +1)
- \tau \sin \phi  \right] \right\}\\[2ex]
\hskip20ex \times \; 
 \exp \left[-\frac{\rho^2}{w_0^2} + \tau \cos \phi  \right]   J_{\vert \ell \vert} \left( \frac{2 \sqrt{2\tau}}{w_0} \rho e^{-\frac{i}{2} \phi }  \right). 
\label{cs2}
\end{array}
\ee
The identities (\ref{iden2}) mean that both, the mode field and the cross-section, change periodically as the beam propagates. Thus, there is a finite spread of the optical power that is reverted periodically at concrete points along the propagation axis.

In turn, function (\ref{cs2}) reveals that the Bessel profile of the coherent states $V_{\ell} (\boldsymbol{r}, z; \xi)$ changes from $I_{\vert \ell \vert}$ to $J_{\vert \ell \vert}$ at the odd multiples of a quarter of the period $2\pi z_R$, no matter the value of the phase $\phi$.

Next, we analyze the propagation properties of the BG coherent states $V_{\ell} (\boldsymbol{r},z; \xi)$  for the real eigenvalue $\xi = -\tau$ ($\phi = -\pi$) and the pure imaginary one $\xi = i$ ($\phi=\frac{\pi}{2}$). The behavior for any other value of $\phi$ is qualitatively equivalent to such cases.

\subsubsection{Behavior for real eigenvalues}

For $\xi = -\tau$, the functions $V_{\ell} (\boldsymbol{r},z; -\tau)$ are defined in Eq.~(\ref{cs1}). The choice $\xi = \tau$ gives rise to an expression quite similar to (\ref{cs1}), but including the modified-Bessel function $I_{\vert \ell \vert}$ instead of $J_{\vert \ell \vert}$. The qualitative behavior of $V_{\ell} (\boldsymbol{r},z; \xi)$ is basically the same for either $\xi=-\tau$ or $\xi =\tau$. Hence, we consider $V_{\ell} (\boldsymbol{r},z; -\tau)$ as representative of the BG coherent states (\ref{cs}) for $\xi \in \mathbb R$.

\begin{figure}[h!]
\centering
\subfloat[][]{\includegraphics[width=.25\textwidth]{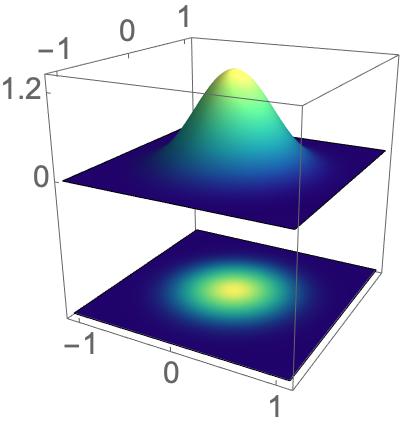}} 
\hskip6ex
\subfloat[][]{\includegraphics[width=.25\textwidth]{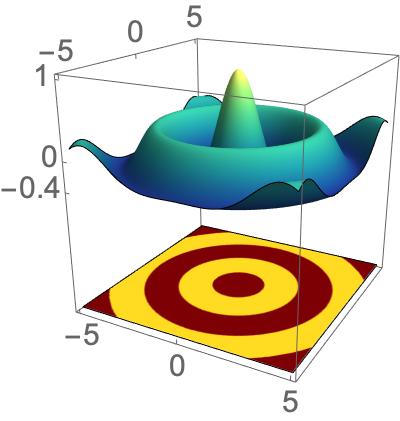}}

\caption{\footnotesize Intensity (\textbf{a}) and phase distribution (\textbf{b}) of the BG coherent state $V_{\ell} (\boldsymbol{r},z; \xi)$ for $\ell=0$ and $\xi = -1/2$. These configurations occur in the transverse planes defined by the points $z_{2q}=2q \pi z_R$. The related profile is characterized by the Bessel function of the first kind $J_0$ depicted at the right, so the phase distribution is radial, with discontinuities located at the zeros of $J_0$ that represent a phase shift $\pm \pi$. The brown zones correspond to positive values of $J_0$ and identify the phase $e^0=1$. In turn, the yellow zones represent negative values of $J_0$, with phase  $e^{i\pi}=-1$.
}
\label{modeJ0}
\end{figure}

$\bullet$ {\bf Initial configuration and periodicity.} Considering $z=0$ as the point of departure, the initial configuration of the beam is recovered at the points $z_{2q} = 2q\pi z_R$, with $q \in \mathbb Z$, see Figure~\ref{propaga}. At any of these points the Bessel function $J_{\vert \ell \vert}$ is real-valued and exhibits a denumerable set of zeros. The latter defines a radial distribution of the phase plane where $J_{\vert \ell \vert}<0$ produces a phase shift $\pi$. In turn, the polar variable $\theta$ sweeps $\vert \ell \vert$ times the interval $[-\pi,\pi)$ in every one of the regions defined by the sign of $J_{\vert \ell \vert}$. 

Figure~\ref{modeJ0} shows the intensity and phase distribution of  $V_0 (\boldsymbol{r},z_{2q}; -\frac12)$. The plots represent the initial configuration of the beam. In this case ($\ell=0$) the polar variable $\theta$ is not present, so the phase plane is covered by a radial distribution of constant phases $0$ and $\pi$, according with the sign of the Bessel function~$J_0$.

\begin{figure}[htb]
\centering
\subfloat[][]{\includegraphics[width=.25\textwidth]{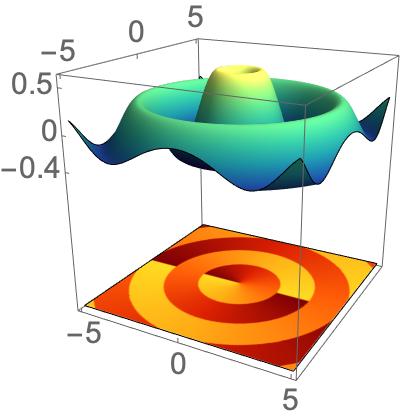}}\hskip4ex
\subfloat[][]{\includegraphics[width=.25\textwidth]{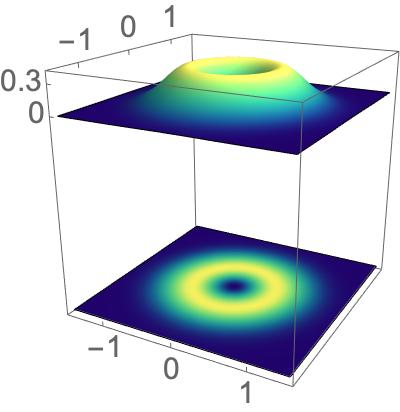}}\hskip4ex
\subfloat[][]{\includegraphics[width=.25\textwidth]{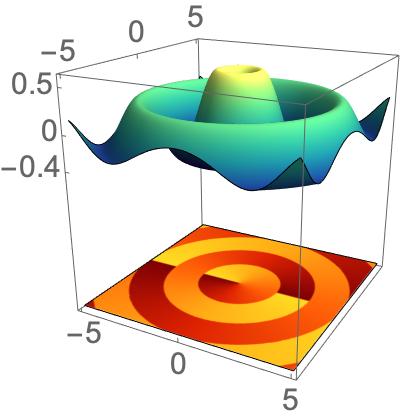}}

\caption{\footnotesize Intensity (\textbf{b}) and phase distribution of the coherent state $V_{\ell} (\boldsymbol{r},z; \xi)$ for $\xi = -1/2$, with $\ell=-1$ (\textbf{a}) and $\ell =1$ (\textbf{c}). These configurations occur in the transverse planes defined by the points $z_{2q}=2q \pi z_R$. The polar phase $\theta$ sweeps $[-\pi, \pi)$ once in each of the radial zones defined by the sign of the Bessel function $J_1$. Brown and yellow color-codes identify the phases $-\pi$ and $\pi$, respectively. The helicity of mode $\ell =-1$ is clockwise while $\ell =1$ is counterclockwise. 
}
\label{modeJ1}
\end{figure}

To get general insights about the behavior of the BG coherent states ({\ref{cs1}) for $\vert \ell \vert \neq 0$, in Figure~\ref{modeJ1} we show the transverse intensity and phase distribution of  $V_1 (\boldsymbol{r},z_{2q}; -\tfrac12)$. The polar phase $\theta$ sweeps $[-\pi, \pi)$ once in each of the radial zones defined by sign of the Bessel function $J_{\vert \ell \vert}$. The helicity of modes $\ell=-1$ and $\ell=1$ is clockwise and counterclockwise, respectively.

\begin{figure}[h!]
\centering
\subfloat[][]{\includegraphics[width=.25\textwidth]{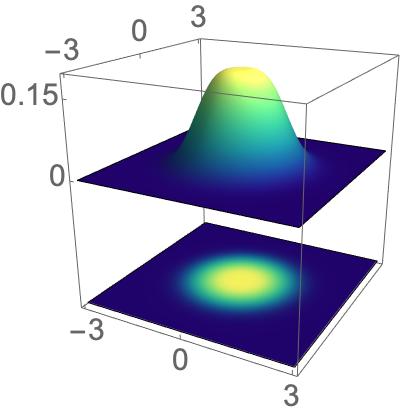}}\hskip6ex
\subfloat[][]{\includegraphics[width=.25\textwidth]{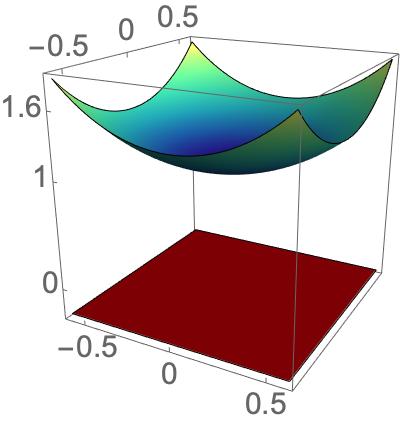}}

\caption{\footnotesize Intensity (\textbf{a}) and phase distribution (\textbf{b}) of the coherent state $V_{\ell} (\boldsymbol{r},z; \xi)$ for $\ell=0$ and $\xi = -1/2$. These configurations occur in the transverse planes defined by the points $z_{2n}=\frac12(4n+1) \pi z_R$. The related profile is characterized by the modified-Bessel function $I_0$, which is positive (this function is depicted at the right). The latter means that, in contrast with $J_0$, the function $I_0$ provides no phase shifts the to coherent state (compare with Figure~\ref{modeJ0}). Thus, the brown color-code refers to a phase $e^0=1$.
}
\label{modeI0}
\end{figure}

$\bullet$ {\bf Self-focusing.}  A second class of interesting points distributted along the propagation axis is defined by the rule $z_q=q\pi z_R$, with $q\in \mathbb Z$. The self-focus of the field is produced twice in a given period $2\pi z_R$, just at the points $z_q$ of the propagation axis, see Figure~\ref{propaga}.

$\bullet$ {\bf Maximum spreading of the beam.} At the points $z_n = \frac12 (2n+1) \pi z_R$, with $n \in \mathbb Z$, the coherent state $V_{\ell} (\boldsymbol{r},z; \xi)$ changes its profile from the Bessel function of the first kind $J_{\vert \ell \vert}$ to the modified-Bessel function $I_{\vert \ell \vert}$. The latter yields the maximum radial uncertainty of the beam. The  intensity distribution is blurred on the transverse plane, and the phase distribution occurs without the radial distribution of the previous cases (the Bessel function $I_{\nu}(u)$ has no zeros along the real axis of the complex $u$-plane if $\nu$ is an integer \cite{Olv10}). Figures~\ref{modeI0} and \ref{modeI1} show the transverse intensity and phase distribution for $\ell=0$ and $\ell=1$, respectively.

\begin{figure}[htb]
\centering

\subfloat[][]{\includegraphics[width=.25\textwidth]{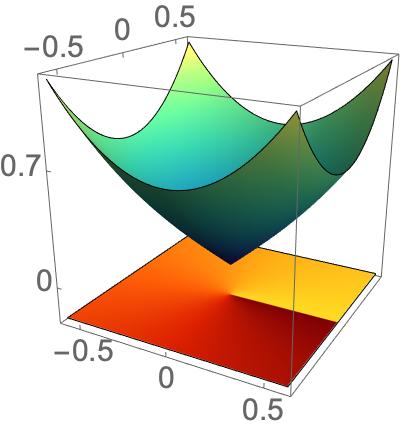}} \hskip4ex
\subfloat[][]{\includegraphics[width=.25\textwidth]{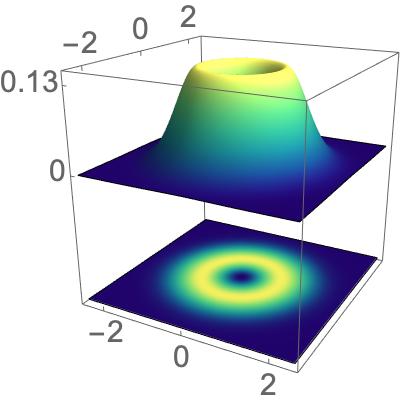}}\hskip4ex
\subfloat[][]{\includegraphics[width=.25\textwidth]{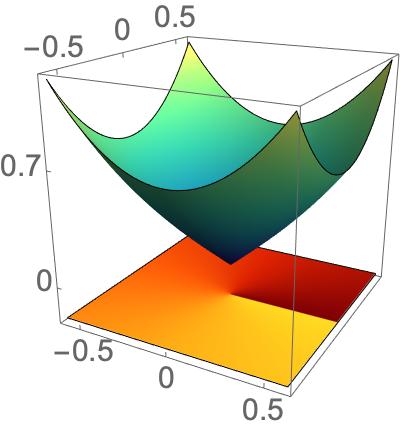}}

\caption{\footnotesize Intensity (\textbf{b}) and phase distribution of the coherent state $V_{\ell} (\boldsymbol{r},z; \xi)$ for $\xi = -1/2$, with $\ell=-1$ (\textbf{a}) and $\ell =1$ (\textbf{c}). These configurations occur in the transverse planes defined by the points $z_{2n}=\frac12(4n+1) \pi z_R$. The polar phase $\theta$ sweeps $[-\pi, \pi)$ once in the transverse plane, brown and yellow color-codes identify the phases $-\pi$ and $\pi$, respectively. The helicity of mode $\ell =-1$ is clockwise while $\ell =1$ is counterclockwise. 
}
\label{modeI1}
\end{figure}

$\bullet$ {\bf Vortices.} At any other point of the propagation axis, the Bessel function appearing in (\ref{cs1}) is complex-valued. Thus, $J_{\vert \ell \vert}$ contributes to the global phase of $V_{\ell} (\boldsymbol{r},z; -\tau)$ with a term that depends on $\rho$ in general. As a consequence, the phase distribution of the initial configuration is distorted such that it exhibits vortices. This is illustrated in Figure~\ref{modeA} for $\ell=1$, the real and imaginary parts of $V_1 (\boldsymbol{r},z; -\tau)$ change sign in different regions of the transverse plane. The phase distribution is therefore characterized by the quotient of such signs.

\begin{figure}[htb]
\centering
\includegraphics[width=.9\textwidth]{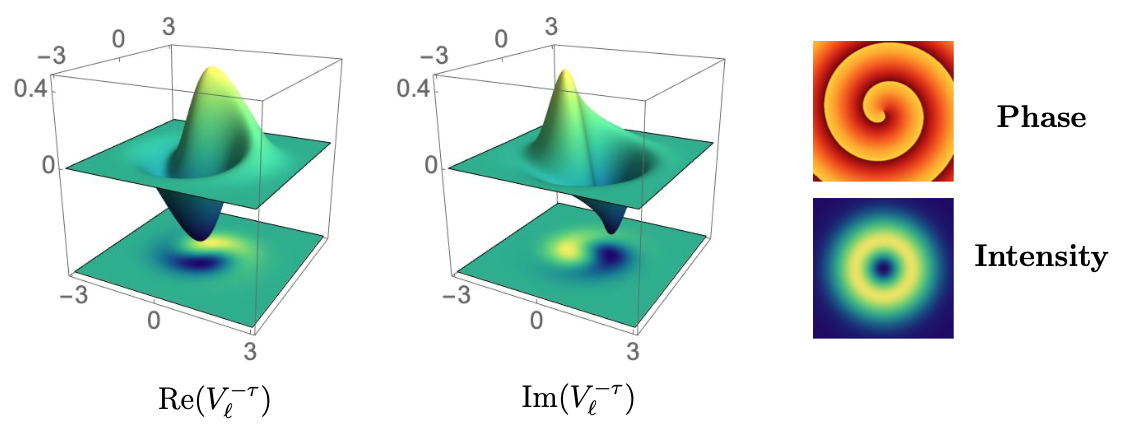}

\caption{\footnotesize Real and imaginary parts of the coherent state $V_{\ell} (\boldsymbol{r},z; \xi)$ for $\ell=1$, $\xi =- 1/2$ and $z=\frac{\pi}{4} z_R$. Their contributions to the polar phase produce the vortex shown in the figure at the right, where it is also shown the field intensity.}
\label{modeA}
\end{figure}

\subsubsection{Behavior for pure imaginary eigenvalues}

Making $\xi=-i \tau$ in (\ref{cs}), the modified-Bessel function $I_{\vert \ell \vert}$ is complex-valued, no matter the value of the propagation variable $z$. As in the previous case, $I_{\vert \ell \vert}$ contributes to the global phase of  $V_{\ell} (\boldsymbol{r},z; -i\tau)$ with a term that depends on $\rho$. See Figures~\ref{modeB} and \ref{modeC}.

\begin{figure}[htb]
\centering
\includegraphics[width=.9\textwidth]{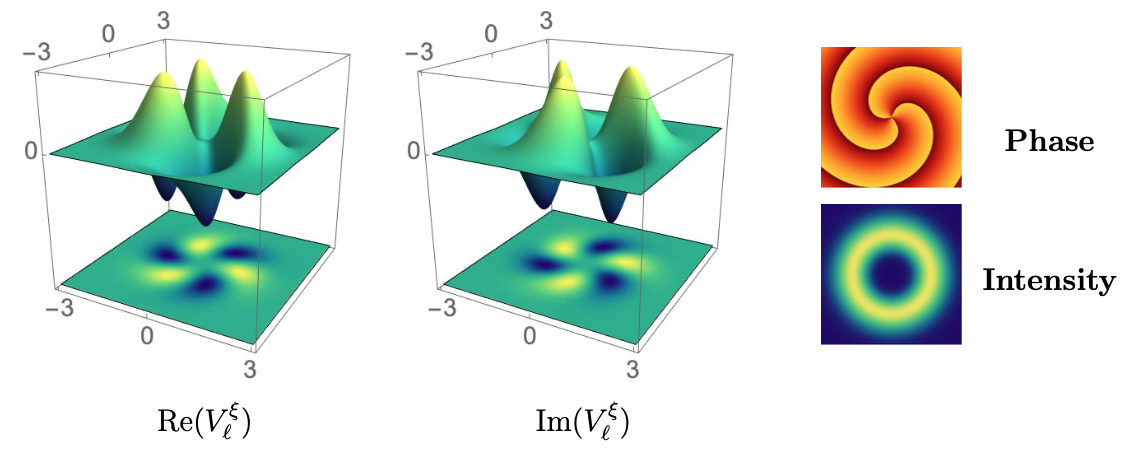}

\caption{\footnotesize Real and imaginary parts of the BG coherent state $V_{\ell} (\boldsymbol{r},z; \xi)$ for $\ell=3$, $\xi = -i/2$ and $z=0$. Their contributions to the polar phase produce the vortex shown in the figure at the right, where it is also shown the field intensity.}
\label{modeB}
\end{figure}

Figure~\ref{modeB} shows the real and imaginary parts of $V_{\ell} (\boldsymbol{r},z; -i\tau)$ at $z=0$, with $\tau = 1/2$. As in the previous case, these functions change sign in different regions of the transverse plane. Note that the number of local maxima is defined by the value of $\ell$, the same holds for the number of local minima (where the functions acquire  negative values). The number of vortices in the phase distribution is also determined by $\ell$. After the beam propagation, at $z= \frac{\pi}{2} z_R$, the distribution of local maxima and minima has changed as well as the configuration of vortices, see Figure~\ref{modeC}.

\begin{figure}[htb]
\centering
\includegraphics[width=.9\textwidth]{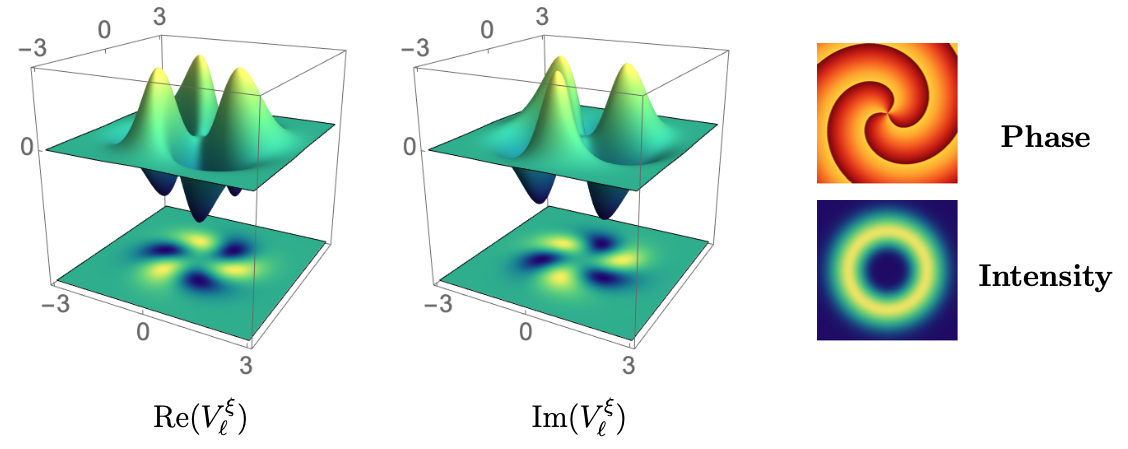}

\caption{\footnotesize Real and imaginary parts of the BG coherent state $V_{\ell} (\boldsymbol{r},z; \xi)$ for $\ell=3$, $\xi = -i/2$ and $z=\frac{\pi}{2} z_R$. Their contributions to the polar phase produce the vortex shown in the figure at the right, where it is also shown the field intensity.}
\label{modeC}
\end{figure}

\section{Discussion}
\label{Sec4}

We have studied the $z$-propagation of electromagnetic waves through weakly inhomogeneous media of parabolic refractive index. After imposing the constant beam-width condition, we focused on the electric field of guided Laguerre-Gauss modes propagating along the $z$-axis with constant width and carrying finite transverse optical power \cite{Cru20}. These exhibit helicity, distinguished by an integer $\ell \neq 0$, which is right-handed for $\ell >0$ and left-handed if $\ell <0$. Modes with no helicity are characterized by $\ell =0$. 

The guided Laguerre-Gauss modes (LG modes for short) form an orthonormal basis for the space of solutions $\mathcal{H}$ of the paraxial wave equation, and can be grouped according to the helicity parameter $\ell$, so they span infinite-dimensional solution subspaces (called hierarchies) $\mathcal{H}_{\ell} \subset \mathcal{H}$ such that $\mathcal{H} = \bigoplus_{\ell} \mathcal{H}_{\ell}$. 

The most striking feature of this decomposition is that the symmetry underlying the hierarchies $\mathcal{H}_{\ell}$ is linked to  the simplest noncompact non-Abelian Lie group \cite{Per86}, denoted by $SU(1,1)$. We have shown that the hierarchies $\mathcal{H}_{\ell}$ are irreducible representation spaces for the Lie algebra $su(1, 1)$, we have also identified the differential form of the generators $\mathcal{L}_{\ell}$, $\mathcal{L}_{\ell}^{\pm}$. 

Surprisingly, the LG modes satisfy the eigenvalue equation of $\mathcal{L}_{\ell}$ with the corresponding propagation constants playing the role of eigenvalues. In turn, $\mathcal{L}_{\ell}^{\pm}$ are ladder operators that connect a certain LG mode to another with eigenvalues stepped by one unit. These results mean that the Lie algebra $su(1,1)$ determines the way the LG modes propagate along the $z$-axis.

Following the notion of coherent states introduced by Barut and Girardello \cite{Bar71}, in each hierarchy $\mathcal{H}_{\ell}$, we have constructed a linear superposition of LG modes that is eigenfunction of $\mathcal{L}_{\ell}^-$ with complex eigenvalue $\xi = \tau e^{-i\phi}$. These generalized coherent states are nothing more than guided Bessel-Gauss modes (BG modes) carrying finite transverse optical power that propagate  with a well-defined optical angular momentum  and self-focus along the $z$-axis. The phase $\phi = \arg (\xi)$ characterizes the profile of the mode (either $J_{\vert \ell \vert}$ or $I_{\vert \ell \vert}$), while the modulus $\tau = \vert \xi \vert$ is responsible for its quality.

We have found that, by setting $\tau$ close to zero, the BG modes so constructed are as Gaussian as the limit $\tau \rightarrow 0$ allows. With this in mind, to evaluate the quality of a signal that is represented by any of these modes, we have analyzed the transverse-spread $\sigma_{\boldsymbol{r}}\sigma_{\boldsymbol{p}}$ of measuring the joint variability of the transverse-position $\boldsymbol{r}$ and the transverse-propagation direction $\boldsymbol{p}$. The Robertson inequality \cite{Rob29} produces a result that depends sinusoidally on the propagation variable $z$ and is such that, the smaller $\tau$, the shorter the difference $\sigma_{\boldsymbol{r}}\sigma_{\boldsymbol{p}} \vert_{\operatorname{max}}- \sigma_{\boldsymbol{r}}\sigma_{\boldsymbol{p}} \vert_{\operatorname{min}}$. In turn, the Schr\"odinger inequality  \cite{Sch30} removes the dependence on  $z$ by providing $\sigma_{\boldsymbol{r}}\sigma_{\boldsymbol{p}} \vert_{\operatorname{min}}$ only. 

Using the above results, to determine how close a BG mode is to an ideal Gaussian beam, we have introduced the quantity $M^2_{\ell} (\tau) = k_0 \sigma_{\boldsymbol{r}}\sigma_{\boldsymbol{p}} \vert_{\operatorname{min}} =2 \sqrt{\langle \mathcal{L}_{\ell} \rangle^2  -\tau^2} \geq 1$, where $\langle \mathcal{L}_{\ell} \rangle$ is the expectation value of $\mathcal{L}_{\ell}$. It is remarkable that $M^2_{\ell} (\tau)$ coincides with the notion of the beam propagation factor (also called beam quality factor) $\mathcal{M}^2$ \cite{Sie90,Sie93}, the most important feature describing the quality of light beams \cite{Sal07,Bel94}. 

It is necessary to emphasize that both, the Robertson and the Schr\"odinger inequalities are commonly used in quantum mechanics but rarely reported in connection with the beam quality factor $\mathcal{M}^2$. Remarkable exceptions are \cite{Dod00}, where it is suggested a link between some invariant quantities and generalized uncertainty relations like the Schr\"odinger one, and \cite{Cru17}, where the $\mathcal{M}^2$ factor of Hermite-Gauss modes is directly related to the Schr\"odinger inequality for $\boldsymbol{r}$ and $\boldsymbol{p}$.

Our expression for $\mathcal{M}^2$ is in complete agreement with the results reported in e.g. \cite{Bor97}, where generic BG beams are studied. The authors of \cite{Bor97} develop a direct calculation of the second-order moments associated to the intensity distributions at the waist plane and in the far field, denoted $\sigma_0$ and $\sigma_{\infty}$ respectively, to write $\mathcal{M}^2 = 2\pi \sigma_0 \sigma_{\infty}$. However, using algebraic techniques, without cumbersome calculations involved, with $M^2_{\ell} (\tau)$ we have introduced a simple and elegant way to evaluate $\mathcal{M}^2$ in terms of $\tau$, the modulus of the complex parameter $\xi$ that characterizes the BG modes as generalized coherent states.

Clearly, not only the profile and the way the BG modes propagate along the $z$-axis, but also their quality is determined by the underlying symmetry of the hierarchy $\mathcal{H}_{\ell}$. This conclusion would not be clear without transferring notions that are quite natural in quantum mechanics to the field of optics. In particular, the construction of ladder operators for LG modes that have a well-defined optical angular momentum allows to create linear superpositions of them, whose coherence properties can be maximized to generate BG modes as generalized coherent states for the Lie algebra $su(1,1)$.  

From the experimental point of view, our BG modes may be produced as the result of the interference of a Gaussian beam with itself \cite{Beu21}, by means of all the available techniques to generate some other BG beams. These include, for instance, the assemblies of circular slits and focusing lenses \cite{Ueh89}, and the use of conical lenses (also called axicons) \cite{Bal21}.

\section{Conclusions}
\label{conclu}

Summarizing some of the most notable coherence properties of the BG modes reported in this work, we have:

\begin{itemize}
\item[$\circ$]
The maximum transverse-spreading of any BG mode is parameterized by $\tau = \vert \xi \vert$. The shorter the value of $\tau$ the better the collimation of the corresponding beam.

\item[$\circ$]
The quality of the BG modes is also parameterized by $\tau$: the shorter the value of $\tau$ the closer the BG modes are to the Gaussian profile.

\item[$\circ$]
The optical angular momentum $\ell \hbar$ spoils the beam quality: poor beam quality results for large $\vert \ell \vert$, no matter how small $\tau $ is.

\item[$\circ$]
The fundamental LG mode is dominant in any superposition intended to build high-quality BG modes. The contribution of the remaining LG modes can be treated as a disturbance (noise) that deviates the BG mode from the ideal Gaussian profile.

\item[$\circ$]
The profile of the BG mode is $I_{\vert \ell \vert}$ for $\phi = \arg(\xi) = 0$, and $J_{\vert \ell \vert}$ for $\phi = -\pi$. 

\item[$\circ$]
No matter the value of $\phi$, the profile of the BG modes changes periodically from $I_{\vert \ell \vert}$ to $J_{\vert \ell \vert}$ as the beam propagates along the $z$-axis.

\item[$\circ$]
The transverse-spreading of the BG modes is always finite and changes periodically from its maximum value to its minimum value at very specific points along the propagation axis.

\item[$\circ$]
At any other point of the propagation axis, the phase distribution of the BG modes exhibit vortices.
\end{itemize}

All these properties are the result of constructing the BG modes as generalized coherent states for the Lie algebra $su(1,1)$, in terms of the irreducible representation spanned by  LG modes that have a well-defined optical angular momentum. 

Like other BG beam models, our generalized coherent states could find applications in optical engineering. For example, to design secure communication protocols \cite{Wan18,Wan23} or improve efficiency in laser writing, such as micro-machining  \cite{Bal21,Beu21} and Bragg grating inscription on materials \cite{Har23}. Higher order BG beams are also efficient for controlling light filamentation in nonlinear materials \cite{Mil21} and for designing optical tweezers \cite{Yan21}. For applications in the quantum domain, since optical angular momentum can be entangled \cite{Kre17},  the BG modes are more suitable for generating and studying entanglement in SPDC processes \cite{McL12}.

It is feasible to use other linear superpositions of LG modes to construct additional coherent states, this time in terms of the Lie group $SU(1,1)$, in the sense proposed by Perelomov \cite{Per86}. In this case, it is well known that there will be several types of coherent states since $SU(1,1)$ has several series of unitary irreducible representations. The analysis of the coherence properties of the resulting modes can be achieved by following the procedure developed in this article. Work is underway in this direction and will be reported elsewhere.

\section*{Acknowledgment}

This research was funded by Consejo Nacional de Ciencia y Tecnolog\'ia (CONACyT, Mexico), grant numbers A1-S-24569 and CF19-304307, and by Instituto Polit\'ecnico Nacional (IPN, Mexico), project SIP20232237. P. J.-M. acknowledges the scholarship support from CONACyT

\appendix
\section{The irreducible representation space of the Lie algebra $su(1,1)$}
\label{ApA}

\renewcommand{\thesection}{A-\arabic{section}}
\setcounter{section}{0}  

\renewcommand{\theequation}{A-\arabic{equation}}
\setcounter{equation}{0}  

From Eqs.~(\ref{phe1}) and (\ref{ugen}) we obtain the differential equation for the stationary LG modes:
\be
\left[ -\frac{w_0^2}2 \left( \frac{\partial^2}{\partial \rho^2} + \frac1{\rho} \frac{\partial}{\partial \rho} - \frac{\ell^2}{\rho^2} \right)+ 2 \frac{\rho^2}{w_0^2} - 2 \beta_{\vert \ell \vert}^p \right] \Phi_{\vert \ell \vert}^p =0.
\label{a1}
\ee
The fundamental solutions of (\ref{a1}) form a complete and orthonormal set in the hierarchy~$\mathcal{H}_{\ell}$, 
\be 
\int_0^\infty \Phi_{\vert \ell \vert}^{\, p *}(\rho)  \Phi_{\vert \ell \vert}^q (\rho) \rho d \rho =  \delta_{p,q}.
\label{orto-s}
\ee
In other words, if $\ell = \operatorname{fixed}$, the stationary LG modes $\Phi_{\vert \ell \vert}^p (\rho)$ form an orthonormal basis for the (stationary) solution subspace 
\[
\mathcal{H}_{\ell} \supseteq \mathcal{V}_{\ell} = \operatorname{span} \{ \Phi_{\vert \ell \vert}^p (\rho); \, \ell = \operatorname{fixed}, \, p=0,1,2,\ldots \}, \quad \ell \in \mathbb Z.
\]
Note that $\mathcal{V}_{\ell}$ and $\mathcal{V}_{-\ell}$ are isomorphic ($\mathcal{V}_{\ell} \simeq\mathcal{V}_{-\ell}$) since their basis elements do not depend on $\operatorname{sgn}(\ell)$. In turn, the stationary subspace $\mathcal{V}_0$ has not a concomitant.

To eliminate the term with the first-order derivative in Eq.~(\ref{a1}), one uses the transformation
\begin{equation} 
\Phi^p_{\vert \ell \vert} (\rho) = \rho^{-1/2} \psi_p (\rho; \ell).
\nonumber
\end{equation}
Then we arrive at the eigenvalue problem
\begin{equation}
H_\ell \, \psi_p    = \varepsilon_p (\ell ) \, \psi_p , \quad p=0,1,2,\ldots, 
\nonumber
\end{equation}
where the eigenvalues $\varepsilon_p (\ell) \equiv 2 \beta_{\vert \ell \vert}^p = 2(\vert \ell \vert +1) + 4p$ are equidistant, with steps of four units $\varepsilon_{p+1}( \ell ) - \varepsilon_p ( \ell ) =4$, and the second-order differential operator
\begin{equation}
H_\ell = \frac{w^2_0}{2} \left[ -\frac{\partial^2}{\partial \rho^2} + \frac{\ell^2 -\frac14}{\rho^2} \right] + \frac{2 \rho^2}{w^2_0}
\label{radHam}
\end{equation}
plays the role of a {\em Hamiltonian}.

From (\ref{orto-s}), it is immediate to verify the orthonormality of the $\psi_p$-functions. Indeed, they satisfy the oscillation theorems of the Sturm theory (see e.g. \cite{Fin76}). Then, one has at hand an additional basis for the subspace $\mathcal{V}_{\ell}$,
\[
\mathcal{V}_{\ell} = \operatorname{span} \{\psi_p(\rho; \ell); \, \ell = \operatorname{fixed}, \, p=0,1,2,\ldots \}, \quad \ell \in \mathbb Z.
\]
The isomorphism $\mathcal{V}_{\ell} \simeq \mathcal{V}_{-\ell}$ implies $\psi_p(\rho;\ell)= \psi_p(\rho; -\ell)$ and $\varepsilon_ p( \ell )  = \varepsilon_p (-\ell)$, so the Hamiltonians $H_{\pm \ell}$ are isospectral (the Hamiltonian $H_{\ell =0}$ has not a concomitant).

The Hamiltonian $H_{\ell}$ may be identified with the radial part of the energy observable of a 2D quantum oscillator in position-representation (see, for example, \cite{Neg00a,Neg00b}). The identification is complete after considering 
\begin{itemize}
\item[(i)] 
The ensemble $\{ \beta_{\vert \ell \vert}^p \}$ is in one-to-one correspondence with the energy spectrum of the 2D quantum oscillator. 
\item[(ii)] 
The {\em square integrability} condition for quantum bound states corresponds to {\em finite transverse optical power} for localized optical beams \cite{Cru17}. 
\end{itemize}
In our case, item (ii) is granted by the orthonormality of the set $\{ \psi_p \}$. 

It is therefore natural to apply quantum mechanical algebraic methods in the study of the LG modes we are dealing with. 

The interest in constructing operators to intertwine the basis elements of the solution space $\mathcal{H}$ of a given dynamical law is not merely mathematical \cite{deL92}. The algebras fulfilled by these operators are connected with the symmetries of both, the space $\mathcal{H}$ itself and the states of the physical system that are represented by the elements of $\mathcal{H}$ \cite{Mil68,Neg00a,Neg00b,Gil94}. Having this in mind, problems arising from eigenvalue equations can be faced in algebraic form, where the symmetries of the system are used to construct solutions in simple and elegant way \cite{deL92,Mil68,Neg00a,Neg00b,Gil94}. An outstanding algebraic approach, known as the factorization method \cite{Inf51}, is commonly used in contemporary quantum mechanics to study a diversity of systems \cite{Mie04}. The main idea is to express a given operator $M_0$ as the product of two additional operators, $M_1$ and $M_2$, so that $M_0= M_1M_2 +\epsilon$, with $\epsilon$ a number called factorization constant. Neither operators $M_{0,1,2}$ are restricted to be self-adjoint nor $\epsilon$ to be real. The factorization operators $M_{1,2}$ provide intertwining relationships  between the eigenfunctions of $M_0$ that may be used to construct the corresponding ladder operators \cite{Neg00a,Neg00b,Mie04,Cru17,Gre19,Cru20}. Here, we address the factorization of the Hamiltonian $H_{\ell}$ to get  ladder operators for the basis elements of $\mathcal{V}_{\ell}$. The results are easily  generalized to $\mathcal{H}_{\ell}$.

For the 2D oscillator-like Hamiltonians (\ref{radHam}), the factorization method yields four different configurations \cite{Neg00a,Neg00b}
\begin{equation}
H_\ell   = a_\ell^+ a_\ell^- + \epsilon_\ell = a_{\ell-1}^- a_{\ell-1}^+ + \epsilon_{\ell-2}
 = b_\ell^+ b_\ell^- - \epsilon_{\ell-2} = b_{\ell +1}^-b_{\ell+1}^+ -\epsilon_\ell,
\label{factor}
\end{equation}
where both, the factorization constant $\epsilon_\ell = 2(\ell + 1)$ and the first-order differential operators
\begin{equation}
a_\ell^\pm = \frac{w_0}{\sqrt2} \left[ \mp \frac{\partial}{\partial \rho} - \frac{\ell+ \frac12}{\rho} \right] + \frac{\sqrt 2}{w_0}\rho, \qquad
b_\ell^\pm = \frac{w_0}{\sqrt2} \left[ \mp \frac{\partial}{\partial \rho} + \frac{\ell - \frac12}{\rho} \right] + \frac{\sqrt 2}{w_0}\rho,
\label{a}
\end{equation}
are labelled by the  orbital angular momentum parameter $\ell \in \mathbb Z$. 

The factorization operators (\ref{a}) satisfy the symmetry relationships $a_{-\ell}^\pm = b_\ell^\pm$, so that $H_{-\ell}$ admits four different factorizations that are equivalent to those given in Eq.~(\ref{factor}), with $\epsilon_{-\ell} = - \epsilon_{\ell-2}$ and $\epsilon_{-\ell -2} = - \epsilon_{\ell}$. Using these symmetries, a given function $\psi_p (\rho; \ell) \in \mathcal{V}_{\ell}$ can be mapped to any other function $\psi_{q} (\rho; l) \in \mathcal{V}_l$, where $q$ and $p$, and $\ell$ and $l$, are different in general. Indeed, from (\ref{factor}), the straightforward calculation gives 
\be
\begin{array}{c}
a_\ell^- H_\ell = \left(H_{\ell+1} + 2\right) a_\ell^-, \qquad 
H_{\ell} a_{\ell}^+= a_{\ell}^+ \left( H_{\ell +1} + 2\right),\\[1ex]
b_\ell^- H_\ell = \left(H_{\ell-1} + 2\right) b_\ell^-, \qquad 
H_{\ell} b_{\ell}^+ = b_{\ell}^+  \left( H_{\ell-1} + 2\right).
\end{array}
\label{inter}
\ee
The bases of $\mathcal{V}_{\ell}$ and $\mathcal{V}_{\ell+1}$ are correlated through the action  of the pair $a_{\ell}^{\pm}$. In turn, the pair $b_{\ell}^{\pm}$ correlates the bases of $\mathcal{V}_{\ell}$ and $\mathcal{V}_{\ell-1}$. Therefore, making $l= \ell+k$, with $k$ an integer, the basis elements of $\mathcal{V}_{\ell}$ can be correlated with those of $\mathcal{V}_l$ by the appropriate combination of $a_{\ell}^{\pm}$ and $b_{\ell}^{\pm}$, including iterations. 

Nevertheless, it must be clear that the factorization operators (\ref{a}) are linear on the entire solution space $\mathcal{V} = \bigoplus_{\ell} \mathcal{V}_{\ell}$, but they are not linear on a given hierarchy $\mathcal{V}_{\ell}$ if the latter is considered isolated from the remaining subspaces of $\mathcal{V}$. In fact, the intertwining relationships (\ref{inter}) correspond to the mappings
\be
a_\ell^-: \mathcal{V}_{\ell} \rightarrow  \mathcal{V}_{\ell +1}, \quad a_\ell^+: \mathcal{V}_{\ell+1} \rightarrow  \mathcal{V}_{\ell}, \quad b_\ell^-: \mathcal{V}_{\ell} \rightarrow  \mathcal{V}_{\ell -1}, \quad b_\ell^+: \mathcal{V}_{\ell-1} \rightarrow  \mathcal{V}_{\ell}.
\label{mappings}
\ee
Then, the domains of $a_{\ell}^-$ and $b_{\ell}^-$ are defined in $\mathcal{V}_{\ell}$ but their ranges are included in $\mathcal{V}_{\ell+1}$ and $\mathcal{V}_{\ell-1}$, respectively. That is, neither $a_{\ell}^-$ nor $b_{\ell}^-$ define an automorphism of $\mathcal{V}_{\ell}$, so they are not linear on $\mathcal{V}_{\ell}$. Similarly, the ranges of $a_{\ell}^+$ and $b_{\ell}^+$ are in $\mathcal{V}_{\ell}$ but their domains are outside such hierarchy. 

We are interested in constructing ladder relationships for the basis elements of $\mathcal{V}_{\ell}$. Thus, we are looking for automorphisms of $\mathcal{V}_{\ell}$. Clearly, none of the factorization operators (\ref{a}) is useful by itself to our purposes. The solution to the problem is in the combined action of $a_{\ell}^{\pm}$ and $b_{\ell}^{\pm}$. 

In the simplest case, if $a_{\ell}^{\pm}$ is applied after its concomitant $a_{\ell}^{\mp}$ (equivalently, if $b_{\ell}^{\pm}$ is applied after $b_{\ell}^{\mp}$ ), the combined action coincides with one of the automorphisms defined by the factorization (\ref{factor}). Of particular interest, the differential ({\em harmonic-number}) operator
\be
\hat n_p = \left\{
\begin{array}{rl}
\frac14 a_{\ell}^+ a_{\ell}^- = \frac14 (H_{\ell}- \epsilon_{\ell}), & \ell \geq 0 \\[2ex]
\frac14 b_{\ell}^+ b_{\ell}^- = \frac14 (H_{-\ell}- \epsilon_{-\ell}), & \ell \leq 0
\end{array}
\right. \, ,
\nonumber
\ee
when acting on $\psi_p(\rho; \ell)$, returns the corresponding number of nodes:
\begin{equation}
\hat n_p  \, \psi_p = p \, \psi_p , \quad \ell = \operatorname{fixed}, \quad p=0,1,2,\ldots 
\nonumber
\end{equation}
To simplify the notation, we have made the dependence of $\hat n_p$ on $\ell$ implicit (it will be made explicit only if necessary). 

At the current stage, we could associate $\hat n_p$ with the photon-number operator of the well-known boson algebra. The higher the number of nodes, the more excited the LG mode. Indeed, as $H_{\ell}$ and $\hat n_p$ commute, the Hamiltonian eigenvalue (propagation constant) $\varepsilon_p ( \ell ) = 2 \beta_{\vert \ell \vert}^p$ is completely determined by the number of nodes (and vice versa). In this sense, according to the value of $p$, the eigenvalues $\varepsilon_p ( \ell )$ define the first -or fundamental- harmonic mode ($p=0$), the second harmonic mode ($p=1$), and so on. 

Using (\ref{mappings}) we see that alternating $a_{\ell}^{\pm}$ and $b_{\ell}^{\pm}$ some less trivial automorphisms are achievable. For example, the products $b_{\ell +1}^- a_{\ell}^-$ and $a_{\ell-1}^- b_{\ell}^-$ provide the same differential operator
\be
A_{\ell}^- = b_{\ell +1}^- a_{\ell}^- =  a_{\ell -1}^- b_{\ell}^- = -H_{\ell} + 2\rho \frac{\partial}{\partial \rho} + 4 \frac{\rho^2}{w_0^2} +  1.
\nonumber
\ee
The Hermitian conjugate of $A_{\ell}^-$ is also useful,
\be
A_{\ell}^+ = a_{\ell}^+ b_{\ell +1}^+ = b_{\ell}^+ a_{\ell -1}^+ = -H_{\ell} - 2\rho \frac{\partial}{\partial \rho}+ 4 \frac{\rho^2}{w_0^2}   - 1; \qquad A_{\ell}^+ = \left( A_{\ell}^- \right)^{\dagger}.
\nonumber
\ee
Indeed, it may be shown that the set
\begin{equation}
L_{\ell}^- = \tfrac14 A_{\ell}^-, \qquad L_{\ell}^+ = \tfrac14 A_{\ell}^+, \qquad 
L_{\ell} = \tfrac14 H_{\ell},
\label{ladder}
\end{equation}
generates the $su(1,1)$ Lie algebra
\be
[ L_{\ell}^-, L_{\ell}^+] = 2L_{\ell}, \qquad [L_{\ell}, L_{\ell}^{\pm} ] = \pm L_{\ell}^{\pm}.
\nonumber
\ee
The feature of (\ref{ladder}) that garners the most attention is that the spectrum of $L_{\ell}$, given by: 
\be
\lambda_p = \frac14 \varepsilon_p ( \ell ) = \frac{\vert \ell \vert}{2} + p + \frac12,
\nonumber
\ee
mimics the  energy distribution (shifted by $\frac12 \vert \ell \vert$) of the one-dimensional harmonic oscillator in quantum mechanics. The analogy is even clearer if we rewrite the Hamiltonian $H_{\ell}$ in terms of the number operator $\hat n_p$. One way of thinking about this property is to consider the hierarchy $\mathcal{V}_{\ell}$ as the  space of stationary states of an oscillator-like system that is represented by the `energy' operator $L_{\ell} = \hat n_p + \frac12 (\ell +1)$. The entire space $\mathcal{V} = \bigoplus_{\ell} \mathcal{V}_{\ell}$ is thus associated with an infinite collection of such oscillator-like systems.

On the other hand, $L_{\ell}^\pm$ are ladder operators for the basis elements of $\mathcal{V}_{\ell}$. Namely, $\psi_{p\pm 1} \propto L_{\ell}^{\pm} \psi_p $ is eigenfunction of $L_{\ell}$ with eigenvalue $\lambda_{p \pm 1} = \lambda_p \pm 1$. Concrete expressions can be derived from the formulae
\be
L_{\ell}^- \psi_0 =0, \quad L_{\ell}^- \psi_p = \sqrt{p (\vert \ell \vert + p)} \, \psi_{p-1}, \quad  L_{\ell}^+ \psi_p = \sqrt{(p+1) (\vert \ell \vert + p+1)} \, \psi_{p+1}.
\label{action}
\ee
From left-equation (\ref{action}) we see that $L_{\ell}^-$ is bounded from below by $\psi_0$, which is free of nodes and belongs to the lowest eigenvalue of the hierarchy. The remaining equations (\ref{action}) allow to reproduce any basis element $\psi_p$ from $\psi_0$.

The hierarchy $\mathcal{V}_{\ell}$ is therefore an irreducible representation space for the Lie algebra  $su(1,1)$. In fact, the eigenvalue $\kappa_{\ell} (\kappa_{\ell}-1)$ of the Casimir operator $C = \tfrac14 (\ell^2 -1) \mathbb I$ yields the {\em Bargmann index} $\kappa_{\ell} =\frac12 (\vert \ell \vert +1)$, with $\mathbb I$ the identity operator in $\mathcal{V}_{\ell}$. This means that the representation is parameterized by a single number $\kappa_{\ell}$, which acquires discrete values $\frac12, 1, \frac32, 2, \ldots$. It is useful to note that the eigenvalues of $L_{\ell}$ acquire now a simpler form $\lambda_p (\ell) = \kappa_{\ell} +p$, with $\kappa_{\ell}$ the lowest eigenvalue of $L_{\ell}$ in the hierarchy $\mathcal{V}_{\ell}$ (see Figure~\ref{energy}).

All algebraic operators that have been defined to act on $\cal V_{\ell} \subseteq \mathcal H_{\ell}$ can be promoted now to operators that act on $\mathcal{H}_{\ell}$. In particular, the factorization operators $a^{\pm}_{\ell}$ and $b^{\pm}_{\ell}$ are respectively replaced by
\begin{equation}
\mathcal{A}_\ell^\pm = e^{\mp i \frac{(z-z_0)}{z_R}}  e^{\mp i \theta}  \frac1{\sqrt{\rho}} \; a_{\ell}^\pm \sqrt{\rho}, \qquad 
\mathcal{B}_\ell^\pm = e^{\mp i \frac{(z-z_0)}{z_R}} e^{\pm i \theta}  \frac1{\sqrt{\rho}} \; b_{\ell}^\pm \sqrt{\rho}.
\nonumber
\end{equation}
The polar and longitudinal and phases of these new operators are closely related to the dynamical properties of the LG modes: they respectively refer to the helicity of the beam and its propagation of along the $z$-axis.

Consistently, the generators of the Lie algebra $su(1,1)$ acquire the form
\begin{equation}
\mathcal{L}_{\ell}^- = e^{i 2\frac{(z-z_0)}{z_R}} \frac1{\sqrt{\rho}} \; L_{\ell}^- \sqrt{\rho}, 
\qquad \mathcal{L}_{\ell}^+ = e^{-i 2\frac{(z-z_0)}{z_R}} \frac1{\sqrt{\rho}} \; L_{\ell}^+ \sqrt{\rho}, \quad 
\mathcal{L}_{\ell} =  \frac1{\sqrt{\rho}} \; L_{\ell} \sqrt{\rho}.
\nonumber
\end{equation}
Thus, the hierarchy $\cal H_{\ell}$ is an irreducible representation space for the $su(1,1)$ Lie algebra.


\end{document}